\documentclass[lettersize,journal]{IEEEtran}
\usepackage{amsmath,amsfonts}
\usepackage{amssymb}
\usepackage{amsthm}
\usepackage{algorithm}
\usepackage{array}
\usepackage[caption=false,font=normalsize,labelfont=sf,textfont=sf]{subfig}
\usepackage{textcomp}
\usepackage{stfloats}
\usepackage{url}
\usepackage{tabularx}
\usepackage{acronym}
\usepackage{verbatim}
\usepackage{graphicx}
\usepackage{tikz}
\usetikzlibrary{arrows.meta,decorations.pathreplacing,calc}
\usepackage{pgfplots}
\pgfplotsset{compat=1.18}
\usepackage{pgfplotstable}
\usepgfplotslibrary{groupplots}
\pgfplotsset{compat=1.18}
\definecolor{normred}{RGB}{213,62,79}
\definecolor{normblue}{RGB}{50,136,189}
\definecolor{normpurple}{RGB}{117,107,177}
\definecolor{normgreen}{RGB}{49,163,84}
\usepackage{graphicx}
\usepackage{cite}
\usepackage{pdfpages}
\usepackage{algpseudocode}
\usepackage{booktabs}
\usepackage{multirow}
\usepackage{siunitx}
\usepackage{hyperref}
\hypersetup{
    colorlinks=true,
    linkcolor=blue,
    filecolor=magenta,      
    urlcolor=cyan,
    pdftitle={Overleaf Example},
    pdfpagemode=FullScreen,
    }

\newtheorem{proposition}{Proposition}
\newtheorem{lemma}{Lemma}

\newtheorem{remark}{Remark}
 
\begin{document}

\acrodef{V2X}{Vehicle-to-Everything}
\acrodef{V2I}{Vehicle-to-Infrastructure}
\acrodef{CAV}{Connected and Autonomous Vehicle}
\acrodef{MILP}{Mixed-Integer Linear Programming}
\acrodef{IM}{Intersection Management}
\acrodef{AIM}{Autonomous Intersection Management}
\acrodef{MIQP}{Mixed-Integer Quadratic Programming}
\acrodef{FCFS}{First Come First Served}
\acrodef{FIFO}{First In First Out}
\acrodef{MARL}{Multi-Agent Reinforcement Learning}
\acrodef{DRL}{Deep Reinforcement Learning}
\acrodef{NMPC}{Nonlinear Model Predictive Control}
\acrodef{DAG}{Directed Acyclic Graph}
\acrodef{PGGO}{Planner-Guided Greedy Ordering}
\acrodef{SOCP}{Second-Order Cone Program}
\acrodef{WCG}{weighted conflict-graph}

\title{Energy- and Time-Efficient Multi-Vehicle Trajectory Planning Using Augmented DAG Search in Configuration Space}
 \author{
 	\parbox{\textwidth}{%
 		\centering
 		Amirreza Akbari$^{1}$, Galina Sidorenko$^{1}$, Johan Thunberg$^{1}$%
 	}%
 	\thanks{$^{1}$Department of Electrical and Information Technology, Lund, Sweden
 		{\tt\small amirreza.akbari@eit.lth.se, galina.sidorenko@eit.lth.se, johan.thunberg@eit.lth.se}}%
    \thanks{This manuscript is currently under review at IEEE Transactions on Intelligent Transportation Systems.}}




\maketitle

\begin{abstract}
Multi-vehicle trajectory planning for \Acfp{CAV} requires balancing energy consumption and travel time while ensuring safety and scalability. While travel-time-optimal methods based on Integer Programming exhibit poor scalability in large-scale multi-vehicle scenarios, existing energy-oriented approaches fail to balance energy consumption and travel time.
We extend a configuration-space framework for multi-vehicle trajectory planning to jointly locally optimize travel time and energy consumption. A weighted energy–travel time objective is combined with an efficient augmented DAG search on two-dimensional decompositions of a high-dimensional configuration space, enabling scalable and collision-free coordination of large numbers of vehicles. Although the proposed framework is applicable to other traffic scenarios, such as roundabouts and vehicle-following, we focus on intersections as a representative and challenging planning problem. To further improve performance for our methodology, we also extend the method with an insertion-ordering heuristic that exploits reduced-space geometric information. Combined with nonlinear model predictive control, the proposed framework is shown to scale to 40 vehicles, outperform MILP-based and convex optimization baselines in dense intersections, and reduce energy consumption at equalized travel times relative to time-optimal planning.
\end{abstract}

\begin{IEEEkeywords}
Autonomous intersection management, connected and automated vehicles, eco-driving, multi-vehicle trajectory planning, configuration space, model predictive control
\end{IEEEkeywords}

\section{Introduction}

\IEEEPARstart{I}{ntersections} are among the most safety-critical and energy-demanding parts of urban road networks. To avoid collisions, conventional signal control repeatedly forces vehicles to decelerate, stop, and re-accelerate, reducing intersection throughput and increasing energy consumption. Frequent acceleration and stop-and-go driving are among the dominant avoidable sources of energy consumption in urban traffic~\cite{BARTH2009400,huang2018eco}. \Acfp{CAV}, enabled by \ac{V2X} communication, enable explicit coordination of crossing sequences and continuous speed profiles, thereby replacing conventional stop-and-go operation~\cite{gholamhosseinian2022comprehensive,hafner2021survey,riostorres2016survey}. The coordination strategy at an intersection directly determines vehicle speed profiles and, consequently, the energy required to traverse it. Energy efficiency should therefore be treated as a primary coordination objective, alongside traffic throughput, rather than as a by-product of time-optimal scheduling.

However, jointly optimizing energy efficiency and travel time while ensuring safety is computationally challenging for large numbers of \acp{CAV}. As a result, existing methods typically sacrifice either solution quality or scalability.
Existing methods typically trade solution quality for computational scalability. Depending on how crossing priorities are assigned, the literature can be broadly divided into \ac{FIFO}-based and scheduling-based approaches. Among the latter, the most direct formulation is centralized mixed-integer optimization, in which binary decision variables encode the relative crossing order of conflicting vehicles~\cite{fayazi2018mixed,7987071}.
Linear delay objectives lead to \acp{MILP}~\cite{fayazi2018mixed}, whereas control-effort or energy objectives lead to \acp{MIQP}~\cite{liu2020high,luo2023computationally}. Although globally optimal for the adopted formulation, both scale exponentially with the number of conflicts, and conservative whole-region reservations further limit intersection capacity.

Hierarchical and decomposition schemes mitigate this cost by separating a discrete scheduling layer from continuous trajectory optimization. A representative energy-relevant instance fixes a globally time-optimal conflict-zone schedule through a binary \ac{MILP} and then solves for an effort- and energy-optimal trajectory realizing that schedule. This relaxes strict \ac{FIFO} ordering to improve utilization~\cite{9611165}. In~\cite{luo2023computationally}, Monte Carlo tree search is used to explore priority sequences and distributed quadratic programs is used for planning the collision-free motions for each sequence. These decomposition methods substantially improve scalability by separating discrete scheduling from continuous trajectory optimization. However, because the crossing order is determined independently of the continuous optimization, its impact on energy consumption is not explicitly considered while the combinatorial scheduling problem still grows with traffic density.

An alternative line of research formulates intersection coordination directly as a continuous optimal control problem with physically motivated energy models. In~\cite{malikopoulos2018decentralized} a decentralized energy-optimal control framework is derived where closed-form solutions for signal-free intersections are provided.
In~\cite{hult2018optimal} energy-optimal coordination with experimental validation is solved and in~\cite{hadjigeorgiou2023realtime} joint optimization of fuel and travel time is done in real time. 

Closest to the present work, \cite{9916133} proposes a hierarchical centralized framework for unsignalized intersections. The method sequentially determines a crossing order and solves a \ac{SOCP} to compute energy--travel-time-optimal trajectories for battery-electric vehicles using a detailed powertrain model that accounts for rolling resistance, aerodynamic drag, motor and transmission losses, and regenerative braking. The authors demonstrate that modest travel-time concessions can yield substantial energy savings. A decentralized robust extension of the framework is presented in~\cite{pan2023hierarchical}. Such methods deliver high-quality, sometimes globally optimal energy-time solutions, but the conic or optimal-control solves become expensive as the number of interacting vehicles grows, which is the scalability barrier the present work targets.

A complementary route to scalability is provided by geometric formulations. In the configuration-space representation of multi-vehicle coordination~\cite{lozano1979algorithm,latombe1991robot}, each vehicle is parameterized by its progress along a fixed path (arc-length parameterized path), while pairwise conflicts are represented as forbidden regions. Coordination is thereby transformed into a path-planning problem, avoiding combinatorial mixed-integer search.
This prioritized decomposition strategy has long been used for multi-robot trajectory coordination~\cite{lavalle2006planning}. Recently, it has been adapted to intersection coordination through two-dimensional decompositions that reduce the high-dimensional configuration-space problem to a sequence of low-complexity graph searches~\cite{akbari20262}.
Geometric planners of this kind, however, are formulated to optimize travel time or coordination length and are therefore agnostic to energy consumption. Consequently, they do not exploit physically motivated energy models that relate vehicle speed and acceleration to fuel consumption and emissions~\cite{ahn2002estimating,RAKHA200449,han2019fundamentals}. Existing methods therefore face a tradeoff: physically motivated energy models do not scale, whereas scalable (geometric) planners do not account for energy consumption. This paper bridges that gap by integrating a physically motivated energy objective into a scalable configuration-space planner and introducing an insertion-ordering heuristic that exploits the planner's reduced-space geometry.

Although applicable to any scenario whose conflicts can be represented as forbidden regions in configuration space, including roundabouts~\cite{naderi2024lanefree}, ramp merging~\cite{le2024stochastic}, and vehicle-following, we focus on intersections as a representative and challenging coordination problem. Intersections concentrate a high density of crossing conflicts within a small area, making them one of the most demanding coordination scenarios. A method that performs well under this conflict density is expected to transfer naturally to traffic scenarios with sparser conflict structures.

\subsection{Contributions}
\label{subsec:contributions}
This paper presents a scalable framework for jointly optimizing energy consumption and travel time in multi-vehicle trajectory planning. The framework combines a physically motivated energy objective with a second-order graph-search method on two-dimensional decompositions of a high-dimensional configuration space, enabling scalable collision-free coordination. 
The main contributions are summarized as follows.

\begin{enumerate}
\item \emph{Joint energy--travel-time objective}.
We formulate multi-vehicle coordination as a configuration-space planning problem that jointly optimizes travel time and a physically motivated expression of energy consumption. The expression combines positive kinetic-energy variations and aerodynamic drag, capturing the acceleration- and speed-dependent components of positive tractive power while assigning no penalty to deceleration. Unlike the Euclidean coordination-length objective used in prior decomposition-based planners~\cite{akbari20262}, the travel-time objective is expressed in the $\ell_\infty$ norm. We prove that this formulation is equivalent to minimizing travel time under individual vehicle speed limits (Proposition~\ref{pro1}).

\item \emph{Energy--travel-time augmented DAG search}.
We develop an incremental planning algorithm that augments the visibility DAG states with incoming segment velocities and velocity-dependent edge costs, accounting for positive kinetic-energy variation and aerodynamic drag. This enables joint optimization of travel time and energy consumption within the graph search trajectory planning. We further show that the resulting algorithm retains a worst-case complexity of $O(Nr^3)$ (Proposition~\ref{prop:complexity}), where $N$ is the number of vehicles and $r$ is the number of conflict regions for the vehicles.

 \item \emph{\ac{PGGO}.} 
Planner-guided insertion ordering. The proposed augmented DAG-search planner computes trajectories for a given vehicle insertion order. As an extension, we propose a deterministic insertion-ordering heuristic (\ac{PGGO}) that exploits the reduced-space geometry generated by the planner to incrementally construct this order while performing trajectory planning. The heuristic combines information on diagonal-blocking conflicts, directional precedence gaps, coupling to already coordinated vehicles, and the conflict pressure induced by the remaining vehicles. We prove that \ac{PGGO} retains the worst-case complexity of $O(Nr^3)$.

\end{enumerate}

The remainder of this paper is organized as follows. Section~\ref{sec:problem_formulation} formulates the coordination problem in configuration space. Section~\ref{sec:energy_aware_incremental} develops the energy-aware incremental coordination algorithm based on augmented DAG-search. Section~\ref{sec:PGGO_ordering} introduces the planner-guided greedy ordering rule. Section~\ref{sec:controller} presents the \ac{NMPC} tracking controller. Section~\ref{sec:numerical_results} reports the numerical results and Section~\ref{sec:conclusion} concludes the paper.

\section{Problem Formulation}
\label{sec:problem_formulation}
\subsection{Configuration-Space Formulation}

We consider $N$ vehicles moving through a controlled intersection. Each vehicle $i\in\{1,\ldots,N\}$ follows a predefined path $\pmb{\phi}_{i}:[0,\bar{s}_i]\rightarrow\mathcal{S}\subset\mathbb{R}^2$, where $s_i$ is the arc-length coordinate along the path and $\bar{s}_i$ is the path length. The joint configuration of all vehicles is $\mathbf{s}= \begin{bmatrix} s_1 & s_2 & \cdots & s_N \end{bmatrix}^{\top}$, and the corresponding configuration space is $\mathcal{C} = [0,\bar{s}_1]\times[0,\bar{s}_2]\times\cdots\times[0,\bar{s}_N]$.

Potential collisions arise when two vehicle paths overlap or intersect in  the physical space $\mathcal{S}$. Such an intersection or overlap of paths is referred to as a conflict, and for such a conflict we introduce so-called conflict regions. For the $k$-th conflict between vehicles $i$ and $j$, let $[\underline{s}_{i}^{\,k},\overline{s}_{i}^{\,k}]$ and $[\underline{s}_{j}^{\,k},\overline{s}_{j}^{\,k}]$ denote the corresponding arc-length intervals for which vehicles $i$ and $j$ occupy the same conflict region. The induced conflict rectangle in the two-dimensional subspace $[0,\bar{s}_i]\times[0,\bar{s}_j]$ is $\mathcal{R}_{ij}^{k} = [\underline{s}_{i}^{\,k},\overline{s}_{i}^{\,k}] \times [\underline{s}_{j}^{\,k},\overline{s}_{j}^{\,k}]$. The corresponding lifted conflict set or forbidden set in the full configuration space is
\begin{equation}
\mathcal{O}_{ij}^{k} = \left\{ \mathbf{s}\in\mathcal{C}: s_i\in[\underline{s}_{i}^{\,k},\overline{s}_{i}^{\,k}],\; s_j\in[\underline{s}_{j}^{\,k},\overline{s}_{j}^{\,k}] \right\}.
\label{eq:lifted_pairwise_obstacle}
\end{equation}
The complete conflict set is then
\begin{equation}
\mathcal{O} = \bigcup_{i<j} \bigcup_k \mathcal{O}_{ij}^{k}.
\label{eq:forbidden_set}
\end{equation}

\subsection{Energy-Aware Objective}
\label{subsec:objective}
The objective is to reduce energy consumption while maintaining collision-free operation and preserving traffic efficiency. To this end, we employ a physically motivated energy term that captures the work associated with positive kinetic-energy variations and aerodynamic drag, rather than modeling a powertrain-specific fuel-consumption map. This term is consistent with microscopic fuel-consumption and emission models based on vehicle speed and acceleration~\cite{ahn2002estimating,RAKHA200449}, as well as eco-driving studies highlighting unnecessary acceleration and stop-and-go driving as dominant sources of fuel consumption~\cite{BARTH2009400,huang2018eco}.
The trade-off between traffic efficiency, i.e., travel time, and energy consumption 
can be expressed as the minimization of the objective below with respect to $T$ and the collection of velocities $\{v_i(\cdot)\}_{i=1}^{N}$.

\begin{equation}
\begin{aligned}
\lambda\,\frac{T}{T_0} &
+ (1-\lambda)\frac{1}{E_0}\sum_{i=1}^{N}
\bigg[ \int_{0}^{T}\operatorname{ReLU}\!\left(m_i v_i(t)\dot v_i(t)\right)\,dt \\
& + \kappa_i \int_{0}^{T} v_i^3(t)\,dt
\bigg].
\end{aligned}
\label{eq:time_objective}
\end{equation}
Here the constants $T_0$ and $E_0$ normalize the travel time and energy terms, respectively, such that $\lambda\in[0,1]$ defines a dimensionless trade-off between traffic efficiency and energy efficiency. The symbol $T$ is the largest travel time among the $N$ vehicles and $v_{i}(t)$ is the velocity of the $i$'th vehicle at time $t$. For notational convenience, the dependence on $t$ is omitted when clear from the context. The symbol $m_{i}$ is the mass of the $i$'th vehicle. The function $\mathrm{ReLU}(x)=\max(0,x)$ is used in this context to penalize positive variations of kinetic energy, and $\kappa_i = \frac{1}{2} \rho C_{d,i}A_i$ is the aerodynamic drag factor of vehicle $i$, where $\rho$ is the air density, $C_{d,i}$ is the drag coefficient, and $A_i$ is the frontal area. The two terms in the energy contribution together act as a surrogate for the positive tractive energy. The first accounts for the inertial component associated with acceleration, while the second accounts for aerodynamic drag. This interpretation is made precise by~\eqref{eq:fuel_decomposition} below, showing that the positive tractive work separates into these two terms plus a constant path-fixed rolling term.

On a flat road, the tractive force of vehicle $i$ comprises inertia, aerodynamic drag, and rolling resistance:
$$
F_i^{\mathrm{tr}} = m_i\dot v_i + \kappa_i v_i^2 + c_{r,i}m_i g,
$$
So the corresponding tractive work $W_i^{\mathrm{tr}}=\int_0^T F_i^{\mathrm{tr}}\,v_i\,dt$ over the interval $[0,T]$ can be decomposed as
\begin{equation}
W_i^{\mathrm{tr}} = K_i + \tilde K_i 
+\kappa_i\!\int_0^T v_i^3\,dt
+ c_{r,i}m_i g\,\bar{s}_i.
\label{eq:fuel_decomposition}
\end{equation}
The inertial term is partitioned into positive and negative contributions: $K_i=\int_{t: \dot v_i >0}\left(m_i v_i(t)\dot v_i(t)\right)\,dt$ 
associated with acceleration phases, and $\tilde K_i=\int_{t: \dot v_i <0}\left(m_i v_i(t)\dot v_i(t)\right)\,dt$, associated with deceleration phases. The rolling term depends only on the path length $\bar{s}_i=\int_0^T v_i\,dt$ and is therefore invariant (i.e., the same) across velocity profiles and coordination orders and is therefore not possible to affect by the planning. The inertial term $K_i$ and the aerodynamic term form the energy-aware component of the objective \eqref{eq:time_objective} capture the dominant energy-consuming mechanisms associated with vehicle acceleration and speed-dependent aerodynamic losses, consistent with the microscopic fuel models in~\cite{ahn2002estimating,RAKHA200449}.

\begin{remark}
The objective \eqref{eq:time_objective} captures the acceleration-related inertial contribution $K_i$ and aerodynamic component of the tractive work while omitting the deceleration-related inertial term $\tilde K_i$ and the path-fixed rolling-resistance term, which is conceptually aligned with recent intersection eco-driving work~\cite{11435876}. At low speeds typical for intersection crossing the acceleration-related inertial term dominates and the aerodynamic term is a small correction~\cite{RAKHA200449}; nevertheless, including it makes the energy term more accurate at higher crossing speeds.
\end{remark}

The physical time formulation in \eqref{eq:time_objective} can be represented in the configuration space $\mathcal{C}$. Let the trajectory $\pmb{\gamma}:[0,1]\rightarrow\mathcal{C}$ describe the joint progress of all vehicles along their predefined paths. Its derivative $\dot{\pmb{\gamma}}(\tau)$ specifies the 
progress rates of the vehicles. We define the normalized vector
\begin{equation}
\beta_i(\tau) = \frac{\dot{\pmb{\gamma}}_i(\tau)} {\left\|\dot{\pmb{\gamma}}(\tau)\right\|_{\infty}}, \qquad i=1,\ldots,N,
\label{eq:beta}
\end{equation}
under the assumption that $\dot{\pmb{\gamma}}(\tau)\neq \mathbf{0}$ on $[0,1]$. Since $\dot{\pmb{\gamma}}(\tau)\geq \mathbf{0}$, this normalization gives $0\leq \beta_i(\tau)\leq 1$ for all vehicles, and at least one vehicle has $\beta_i(\tau)=1$. $\beta_i(\tau)$ represents the relative velocity contribution of vehicle $i$. The physical reference velocity along the path is obtained by multiplying this relative velocity factor by the maximum admissible velocity:
\begin{equation}
v_{i}^{\mathrm{ref}}(\tau) = \beta_{i}(\tau)v_{\max}.
\label{eq:velocity_reference}
\end{equation}

Using this velocity representation, we formulate the following configuration-space optimization problem 
\begin{equation}
\begin{aligned}
&\min_{\pmb{\gamma}} \quad \lambda \frac{1}{L_0} \int_{0}^{1}
\left\|\dot{\pmb{\gamma}}(\tau)\right\|_\infty d\tau  \\
& \quad + (1-\lambda)\frac{1}{E_0}\sum_{i=1}^{N}\Bigg[
\int_{0}^{1}\mathrm{ReLU}\!\left(\frac{d}{d\tau}\!\left[\tfrac12 m_i(\beta_i(\tau)v_{\max})^2\right]\right)d\tau \\
& \qquad\qquad\quad + \kappa_i v_{\max}^2 \int_0^1 \beta_i^3(\tau)\,\|\dot{\pmb{\gamma}}(\tau)\|_\infty\,d\tau \Bigg] \\
& \mathrm{s.t.}\quad\pmb{\gamma}(\tau)\in\mathcal{C}\setminus\mathcal{O}, \quad \dot{\pmb{\gamma}}(\tau)\geq\mathbf{0}, \quad \pmb{\gamma}(0)=\mathbf{0}, \quad \pmb{\gamma}(1)=\bar{\mathbf{s}},
\end{aligned}
\label{eq:energy_objective}
\end{equation}
where $L_{0}=\|\bar{\mathbf{s}}\|_\infty$ for $\bar{\mathbf{s}} = \begin{bmatrix} \bar{s}_{1} & \bar{s}_{2} & \cdots & \bar{s}_{N} \end{bmatrix}^{\top}$,
 $E_0=\frac{1}{2}\sum_{i=1}^{N}m_i(v_{\max})^2 + \sum_i \kappa_iv_{max}^2 \bar s_i$ is the nominal kinetic energy required to accelerate all vehicles from rest to $v_{\max}$. The first term in \eqref{eq:energy_objective} is equivalent to the minimum travel time under identical individual speed limits as shown in Proposition~\ref{pro1}. 

The second term evaluates the positive kinetic-energy variation and the aerodynamic dissipation using the reference velocity \eqref{eq:velocity_reference}; the aerodynamic contribution is the configuration-space image of $\kappa_i \int_0^T v_i^3 dt$ under the time-optimal parametrization of Proposition~\ref{pro1}.

\begin{proposition}\label{pro1}
Let $\pmb{\gamma}:[0,1]\rightarrow \mathcal{C}$ be absolutely continuous and monotone, i.e.,
$\dot{\gamma}_i(\tau)\geq 0$ almost everywhere 
for $i\in\{1,\ldots,N\}$. Assume each vehicle $i$ of the $N$ vehicles satisfies the individual speed constraint $0\leq \dot{s}_i(t)\leq v_{\max}$. Then the minimum travel time required to execute the geometric path $\pmb{\gamma}$ is
\begin{equation}\label{eq:time}
   T^\star(\pmb{\gamma}) = \frac{1}{v_{\max}} \int_0^1 \|\dot{\pmb{\gamma}}(\tau)\|_\infty d\tau.
\end{equation}
\end{proposition}
\begin{proof}
Let 
$\pmb{s}(t)=\pmb{\gamma}(\tau(t))$, where $\tau(t)$ is a nondecreasing time parametrization satisfying $\tau(0)=0$ and $\tau(T)=1$. Almost everywhere, $\dot{\pmb{s}}(t) = \dot{\pmb{\gamma}}(\tau(t))\dot{\tau}(t)$.
The individual speed constraints imply, for every vehicle $i$,
$\dot s_i(t) = \dot{\gamma}_i(\tau(t))\dot{\tau}(t) \leq v_{\max}$. Therefore, $\|\dot{\pmb{\gamma}}(\tau(t))\|_\infty \dot{\tau}(t) \leq v_{\max}.$
Equivalently, wherever $\dot{\tau}(t)>0$,
\begin{equation}
    \frac{dt}{d\tau} \geq \frac{\|\dot{\pmb{\gamma}}(\tau)\|_\infty}{v_{\max}}.
\end{equation}
Integrating over $[0,1]$ gives the lower bound
\begin{equation}\label{eq:lower_bound:1}
    T \geq \frac{1}{v_{\max}} \int_0^1 \|\dot{\pmb{\gamma}}(\tau)\|_\infty d\tau.
\end{equation}
It remains to show that the lower bound \eqref{eq:lower_bound:1} is attainable, i.e., that $\dot s_i(t) \geq 0$ is ensured for all $i$. Let
\begin{equation}
    \frac{dt}{d\tau} = \frac{\|\dot{\pmb{\gamma}}(\tau)\|_\infty}{v_{\max}}.
\end{equation}
Then, almost everywhere, $\tau$, $\dot{s}_i(t)= v_{\max} \frac{\dot{\gamma}_i(\tau)} {\|\dot{\pmb{\gamma}}(\tau)\|_\infty}$.
Because $\pmb{\gamma}$ is monotone and $\dot{\gamma}_i(\tau)\leq \|\dot{\pmb{\gamma}}(\tau)\|_\infty$, it follows that
$0\leq \dot{s}_i(t)\leq v_{\max}$. Hence the lower bound is attainable, which proves the result.
\end{proof}

The two-dimensional decomposition-based formulation in~\cite{akbari20262} minimizes the Euclidean configuration-space length. In contrast, the proposed formulation here uses the infinity norm because the maximum progress component determines the admissible time scaling under individual vehicle speed limits. The standard norm relation below is useful for comparing the Euclidean and infinity-norm coordination lengths~\cite{golub2013matrix}. While the $L_2$-norm corresponds to aggregate progress in configuration space, the $L_\infty$-norm captures the largest progress component.

\begin{lemma}\label{lemma1}
Let $\pmb{\gamma}:[0,1]\rightarrow\mathcal{C}$ be an absolutely continuous coordination path. Define
$$
L_p(\pmb{\gamma})= \int_0^1 \|\dot{\pmb{\gamma}}(\tau)\|_p\,d\tau, \qquad p\in\{2,\infty\}.
$$
Then
$$
L_\infty(\pmb{\gamma}) \leq L_2(\pmb{\gamma}) \leq \sqrt{N}\,L_\infty(\pmb{\gamma}).
$$
Moreover, if on a trajectory segment 
$[\tau_1, \tau_2]$
exactly $k\leq N$ vehicles move with the same (positive) progress rate, while the remaining vehicles remain stationary, then the Euclidean segment cost is $\sqrt{k}$ times the corresponding infinity-norm segment
cost. 
\end{lemma}

The remaining challenge lies in addressing the optimization problem \eqref{eq:energy_objective}. In the next section, we tackle this problem by combining an incremental two-dimensional decomposition of the configuration space with an augmented second order graph-search procedure that incorporates the energy-aware objective directly into the planning step.

\section{Energy-Aware Incremental Coordination}
\label{sec:energy_aware_incremental}

This section presents an efficient method for  \eqref{eq:energy_objective} for a given vehicle order $\pi$. 
The method uses the two-dimensional decomposition principle of~\cite{akbari20262}, but differs in the graph-search objective. Instead of minimizing only geometric path length, the proposed search augments the graph state with the incoming segment velocity and assigns each edge a cost that combines path length with energy variation. Thus, energy-aware velocity shaping is incorporated directly into the trajectory planning. 

Together, sections \ref{sec:energy_aware_incremental} and \ref{sec:PGGO_ordering} constitute the complete solution framework for problem \eqref{eq:energy_objective}, yielding a suboptimal solution.
It is important to distinguish the vehicle insertion order $\pi$ from the scheduling order of the vehicles through the intersection. The insertion order $\pi$ determines the sequence in which vehicles are incorporated into the incremental planning procedure, whereas the scheduling order is determined by the resulting coordination trajectory obtained as the output of the proposed solution.
The method in Section \ref{sec:energy_aware_incremental} assumes a given insertion order $\pi$, whereas the method, subsequently presented in Section \ref{sec:PGGO_ordering} further down, presents an extension of this method, where  for selecting a suitable insertion order $\pi$ is obtained jointly together with the trajectory.

\subsection{Incremental Algorithm}
\label{subsec:incremental_construction}

Let $\pi=(i_1,i_2,\ldots,i_N)$ denote the order in which vehicles are incorporated, i.e., the insertion order. The proposed method constructs the full  coordination trajectory by solving a sequence of two-dimensional reduced planning problems. The first reduced problem considers vehicles $i_1$ and $i_2$ in $\mathcal{C}^{1} = [0,\bar{s}_{i_1}] \times [0,\bar{s}_{i_2}]$. The pairwise conflict regions between these two vehicles are projected onto $\mathcal{C}^{1}$, where they form rectangular forbidden regions (conflict regions). Solving the reduced problem gives a collision-free monotone path $\pmb{\gamma}^{1}:[0,1]\rightarrow\mathcal{C}^{1}$. Let $L^{1}=\|\pmb{\gamma}^{1}\|_\infty$ denote its $L_\infty$ arc length.

Suppose that after step $n-1$, a feasible trajectory $\pmb{\gamma}^{n-1}$ has been obtained for vehicles $\{i_1,\ldots,i_n\}$. Let $L^{n-1}$ denote the cumulative length parameter for $\pmb{\gamma}^{n-1}$. To add vehicle $i_{n+1}$, we define the next reduced space as $\mathcal{C}^{n} = [0,\bar{s}_{i_{n+1}}] \times [0,L^{n-1}]$. The first coordinate represents the progress of the newly added vehicle, while the second coordinate represents progress along the previously constructed trajectory $\pmb{\gamma}^{n-1}$. Pairwise conflicts between vehicle $i_{n+1}$ and the already coordinated vehicles are projected onto $\mathcal{C}^{n}$ as rectangular forbidden regions.

This procedure is continued until all $N$ vehicles have been incorporated, after which the final reduced trajectory $\pmb{\gamma}^{N-1}$ is recursively lifted through the intermediate trajectories to obtain the full coordination trajectory $\pmb{\gamma}$. The algorithm is incremental because vehicles are introduced sequentially, and each addition requires solving only a two-dimensional reduced planning problem involving the new vehicle and the aggregate trajectory of the previously coordinated group.

\subsection{Solving Two-Dimensional Problems}
Each reduced problem is solved using a state-augmented monotone visibility \ac{DAG}. Consider a reduced space $\mathcal{C}'=[0,\bar{s}_x]\times[0,\bar{s}_y]$ with rectangular forbidden regions collected in $\mathcal{R}'$. The geometric vertex set is $\mathcal{V} = \{\mathbf{p}_0,\mathbf{p}_f\} \cup \mathcal{V}_{\mathrm{obs}}$, where $\mathbf{p}_0=[0,0]^{\top}$, $\mathbf{p}_f=[\bar{s}_x,\bar{s}_y]^{\top}$, and $\mathcal{V}_{\mathrm{obs}}$ contains the corners of the rectangular forbidden regions. This construction follows classical visibility-graph methods for planning around polygonal obstacles \cite{lozano1979algorithm,latombe1991robot}. A directed edge $e=(\mathbf{p}_a,\mathbf{p}_b)$ is included if the line segment from $\mathbf{p}_a$ to $\mathbf{p}_b$ does not intersect the interior of any obstacle in $\mathcal{R}'$ and satisfies $\mathbf{p}_b-\mathbf{p}_a\geq \mathbf{0}$. This condition enforces monotonic progress in both reduced coordinates. Since every admissible edge increases at least one coordinate and never decreases either coordinate, the resulting directed graph is acyclic and can be solved by relaxing vertices in topological order \cite{cormen2009introduction}.

The proposed graph search is not purely geometric because the cost of an edge depends on the velocity of the preceding edge. Such transition-dependent costs can be handled by augmenting the graph state, similarly to shortest-path formulations with turn or transfer penalties \cite{winter2002modeling,lewis2020algorithms}. Accordingly, the search state is
defined as
\begin{equation}
z=(\mathbf{p},\mathbf{v}_{\mathrm{in}}),
\label{eq:augmented_state}
\end{equation}
where $\mathbf{p}\in\mathcal{V}$ is the current geometric vertex and $\mathbf{v}_{in}$ is the nominal velocity of the incoming edge. Hence, the same geometric vertex may correspond to multiple augmented states with different incoming velocities.

Since the final physical velocities are obtained only after the full coordination trajectory is reconstructed, each admissible edge in the two-dimensional reduced space is assigned a nominal velocity for evaluating transition costs. For an admissible edge $e=(\mathbf{p}_a,\mathbf{p}_b)$ with $\Delta\mathbf{p}_e=\mathbf{p}_b-\mathbf{p}_a \in \mathbb{R}_{\ge 0}^{2} \setminus \{\mathbf{0}\}$, we define $\mathbf{v}_e = v_{\max}\frac{\Delta\mathbf{p}_e}{\|\Delta\mathbf{p}_e\|_\infty}$. This normalization preserves the edge progress ratio and saturates the largest reduced-coordinate velocity at $v_{\max}$. The nominal velocity is used only for graph-cost evaluation and is not the final vehicle velocity.

At a graph node in the DAG, 
let $\mathbf{v}_{\mathrm{in}}=[v_{\mathrm{in}, 1},v_{\mathrm{in}, 2}]^{\top}$ denote the incoming nominal velocity and 
Let $\mathbf{v}_{e}=[v_{e,1},v_{e,2}]^{\top}$ denote the velocity associated with a candidate outgoing edge $e$ in the reduced two-dimensional space $\mathcal{C}'$. For each coordinate $r \in \{1,2\}$ in the reduced space $\mathcal{C}'$, referred to as a reduced coordinate, we define the effective mass, $m^{\mathrm{eff}}_{r}$, and the effective aerodynamic coefficient, $\kappa^{\mathrm{eff}}_{r}$, in the following manner. If coordinate $r$ corresponds to a single vehicle $i$, then $m^{\mathrm{eff}}_{r}=m_i$ and $\kappa^{\mathrm{eff}}_{r} = \kappa_i$. If coordinate $r$ corresponds to a previously coordinated group of vehicles $\mathcal{G}_r \subset \{1,2, \ldots, N\},$ then the effective coefficients are evaluated locally as

\begin{equation}
    m^{\mathrm{eff}}_{r} = \sum_{i\in\mathcal{G}_r} m_i\beta_i^2,
\qquad
\kappa^{\mathrm{eff}}_{r} = \sum_{i\in\mathcal{G}_r}\kappa_i\beta_i^3,
\label{eq:effective_coeffs}
\end{equation}
where $\beta_i$ is the relative velocity factor of vehicle $i$ on the corresponding segment of the previously constructed group trajectory \eqref{eq:beta}. 

The positive kinetic-energy increment associated with taking edge $e$ is computed relative to the incoming nominal velocity as
\begin{equation}
\Delta E^{+}_{e} = \sum_{r=1}^{2} \mathrm{ReLU} \left(\frac{1}{2}m^{\mathrm{eff}}_{r}(v_{e, r})^{2} - \frac{1}{2}m^{\mathrm{eff}}_{r}(v_{\mathrm{in}, r})^{2}
\right).
\label{eq:edge_energy_penalty}
\end{equation}
The aerodynamic work along edge $e$ is obtained from the nominal edge velocity. Since $\int v^3\,dt=v^2\!\int v\,dt$ at constant nominal speed, the per-edge aerodynamic penalty is calculated as
\begin{equation}
\Delta E_e^{\mathrm{aero}} = \sum_{r=1}^{2}\kappa_r^{\mathrm{eff}}\,v_{e,r}^{2}\,\Delta p_{e,r},
\label{eq:edge_aero_penalty}
\end{equation}
where $\Delta p_{e,r}$ is the progress with respect to coordinate $r$ along the edge, $\Delta\mathbf{p}_e=[\Delta p_{e,1},\Delta p_{e,2}]^{\top}$.
 Unlike the kinetic penalty, $\Delta E_e^{\mathrm{aero}}$ does not depend on the incoming velocity and accrues on every edge with positive progress.

\begin{proposition}
\label{prop:effective_mass}
Suppose $r$ is a reduced coordinate corresponding to a coordinated group $\mathcal{G}_r$ and suppose each vehicle $i\in\mathcal{G}_r$ moves with $v_i=\beta_i u_r$, where $u_r$ is the reduced-coordinate velocity $v_{max}\frac{\Delta \mathbf{p}_{e,r}}{\|\Delta \mathbf{p}_e\|_\infty}$. Then the group kinetic energy and aerodynamic resistance are represented by the effective coefficients defined in \eqref{eq:effective_coeffs}, in the sense that
$\sum_{i\in\mathcal{G}_r}\tfrac12 m_i v_i^2=\tfrac12 m_r^{\mathrm{eff}}u_r^2$ and
$\sum_{i\in\mathcal{G}_r}\kappa_i v_i^3=\kappa_r^{\mathrm{eff}}u_r^3$.
\end{proposition}
\begin{proof}
Substituting $v_i=\beta_i u_r$, yields
$$
\sum_{i\in\mathcal{G}_r}\tfrac12 m_i v_i^2=\tfrac12\Big(\sum_{i\in\mathcal{G}_r}m_i\beta_i^2\Big)u_r^2,
\quad
\sum_{i\in\mathcal{G}_r}\kappa_i v_i^3=\Big(\sum_{i\in\mathcal{G}_r}\kappa_i\beta_i^3\Big)u_r^3,
$$
which gives the effective coefficients in \eqref{eq:effective_coeffs}.
\end{proof}

\eqref{eq:edge_energy_penalty} and \eqref{eq:edge_aero_penalty} are discrete counterparts of the kinetic and aerodynamic terms in \eqref{eq:energy_objective}. The resulting edge cost is
\begin{equation}
c(e) = \lambda \frac{\|\Delta\mathbf{p}_e\|_\infty}{L_0'} + (1-\lambda)\frac{\Delta E_e^{+}+\Delta E_e^{\mathrm{aero}}}{E_0'},
\label{eq:augmented_edge_cost}
\end{equation}
where $L_0^{\prime}$ and $E_0^{\prime}$ are 2-D space normalization constants. 
For fixed vehicle order $\pi$, the complete procedure of the proposed approach is summarized in Algorithm~\ref{alg:incremental_state_augmented_dag}.

\begin{algorithm}[t]
\caption{Incremental State-Augmented DAG Coordination}
\label{alg:incremental_state_augmented_dag}
\begin{algorithmic}
\Require Vehicle order $\pi=(i_1,\ldots,i_N)$, paths $\{\pmb{\phi}_i\}$, conflict set $\mathcal{O}$, masses $\{m_i\}$, aerodynamic factors $\{\kappa_i\}$, $v_{\max}$, $\lambda$
\Ensure Coordination trajectory $\pmb{\gamma}$
\State $\mathcal{C}^{1}\leftarrow[0,\bar{s}_{i_1}]\times[0,\bar{s}_{i_2}]$
\State $\mathcal{O}^{1}\leftarrow$ project conflicts between $i_1$ and $i_2$ onto $\mathcal{C}^{1}$
\State $\pmb{\gamma}^{1}\leftarrow
\textsc{EAGS}(\mathcal{C}^{1},\mathcal{O}^{1},\{m_i\},\{\kappa_i\},v_{\max},\lambda)$
\For{$n=2$ to $N-1$}
    \State $L^{n-1}\leftarrow L_\infty(\pmb{\gamma}^{n-1})$
    \State $\mathcal{C}^{n}\leftarrow[0,\bar{s}_{i_{n+1}}]\times[0,L^{n-1}]$
    \State $\mathcal{O}^{n}\leftarrow$ project conflicts between $i_{n+1}$ and $\{i_1,\ldots,i_n\}$ onto $\mathcal{C}^{n}$
    \State $\widehat{\pmb{\gamma}}^{n}\leftarrow
    \textsc{EAGS}(\mathcal{C}^{n},\mathcal{O}^{n},\{m_i\},\{\kappa_i\},v_{\max},\lambda,\pmb{\gamma}^{n-1})$
    \State $\pmb{\gamma}^{n}\leftarrow
    \textsc{Lift}(\widehat{\pmb{\gamma}}^{n},\pmb{\gamma}^{n-1})$
\EndFor
\State \Return $\pmb{\gamma}=\pmb{\gamma}^{N-1}$
\end{algorithmic}
\end{algorithm}

A key function in Algorithm~\ref{alg:incremental_state_augmented_dag} is $\textsc{EAGS}$ (Energy-aware graph search), which constructs the monotone visibility DAG, augments each state with the incoming nominal velocity, and assigns the edge cost in \eqref{eq:augmented_edge_cost}, where the group-axis effective coefficients $m_r^{\mathrm{eff}}$ and $\kappa_r^{\mathrm{eff}}$ in \eqref{eq:effective_coeffs} are evaluated from the relative velocity factors $\beta_i$ along the group trajectory $\pmb{\gamma}^{n-1}$. For the first reduced problem in the space $\mathcal{C}^1$, both dimensions correspond to single vehicles, so the effective coefficients reduce to the individual $m_{i}$ and $\kappa_{i}$. The resulting DAG is then solved by topological relaxation to obtain the minimum-cost monotone path $\widehat{\pmb{\gamma}}^{n}$. The function $\textsc{Lift}$ maps the reduced path $\widehat{\pmb{\gamma}}^{n}$ to the higher-dimensional coordination space using the second coordinate of $\widehat{\pmb{\gamma}}^{n}$ as the arc-length parameter along $\pmb{\gamma}^{n-1}$. The feasibility of the lifted trajectory follows from the same argument as in~\cite{akbari20262}. The proposed energy-aware extension modifies only the edge cost; it does not change the admissible edge set, monotonicity constraint, or lifting operation. Hence, the collision-free feasibility of the incremental construction is preserved.

\begin{proposition}
\label{prop:complexity}
The worst-case time complexity of the proposed energy-aware incremental coordination algorithm is $O(Nr^3)$, where $r$ denotes the maximum number of projected rectangular forbidden regions in any reduced problem.
\end{proposition}

\begin{proof}
In the original incremental decomposition~\cite{akbari20262}, each reduced visibility graph has $O(r)$ vertices and $O(r^2)$ monotone edges. In the proposed energy-aware extension, the cost of an edge also depends on the incoming nominal velocity. Therefore, each geometric vertex must be augmented by its incoming edge, resulting in $O(r^2)$ augmented states. In the worst case, each augmented state can transition through $O(r)$ outgoing monotone edges, giving $O(r^3)$ augmented transitions per reduced problem. Since the augmented graph remains acyclic, topological relaxation solves each reduced problem in $O(r^3)$ time. The incremental algorithm solves $N-1$ such reduced problems, which gives the overall complexity $O(Nr^3)$.
\end{proof}
Compared with the geometric incremental search in~\cite{akbari20262}, the energy term increases the worst-case 2-D search complexity from $O(r^2)$ to $O(r^3)$. However, the increase of computational cost is moderate in practice because the number of projected conflict rectangles in each reduced 2-D problem is typically small, and the corresponding visibility graphs are usually sparse.

\section{Planner-Guided Greedy Ordering}
\label{sec:PGGO_ordering}

The incremental coordination presented in  Section~\ref{sec:energy_aware_incremental} requires an insertion order $\pi=(i_1,\ldots,i_N)$, which strongly affects both travel time and energy. Exhaustive search over the $N!$ permutations of this order is intractable, random orders give poor quality, and the local search for random a random initialization of~\cite{akbari20262} explores only a limited neighborhood of the initialization. To overcome this, we here propose \acf{PGGO}, a deterministic rule that builds $\pi$ one vehicle at a time from planner-relevant geometric criteria and graph-based criteria. This is done at negligible computational cost relative to that of Algorithm~\ref{alg:incremental_state_augmented_dag}.

Let $\mathcal A\subseteq\{1,\ldots,N\}$. 
The goal is to construct the permutation $\pi$ sequentially by selecting $i_1$, then $i_2$, and so on. Initially, $\pi$ is the empty vector. At iteration $k$ of this process, \acf{PGGO} selects the next vehicle $i_k$ and then updates $\pi$ according to
$\pi \gets (\pi, i_k)$. Let 
  $\mathcal G = \{i_1,\ldots,i_{k-1}\}\subseteq \mathcal A$, which comprises the group of vehicles already inserted. Let $\mathcal U=\mathcal A\setminus\mathcal G$, which comprises the remaining candidates. The \acf{PGGO}-method  selects $i_k$ as the $\tilde i$ in $\mathcal U$ that minimizes so-called the insertion score $\Psi(\tilde i \mid\mathcal G)$ defined below. 

\begin{equation}
\begin{aligned}
\Psi&(\tilde i\mid \mathcal G)
= \omega_d D(\tilde i,\mathcal G)-\omega_g\Gamma(\tilde i,\mathcal G)
   -\omega_{v} V(\tilde i,\mathcal G)-\omega_\xi\Xi(\tilde i,\mathcal G) \\
& +(1-r_n)\Big[\omega_\ell\,\ell_{{\tilde i}}      
   -\omega_f\big(\theta_F F({\tilde i},\mathcal U\setminus\{{\tilde i}\})+(1-\theta_F)\hat c_{\tilde i}\big)\Big] \\
& -\omega_c\, r_n\, C({\tilde i},\mathcal G)-\omega_m \hat{m}_{\tilde i}.
\end{aligned}
\label{eq:pggo_selection}
\end{equation}
where all $\omega$ coefficients are nonnegative parameters. The terms encode three sources of information: the reduced-space conflict geometry between the candidate and the group ($D,\Gamma,V,\Xi$), the candidate's structure in a weighted conflict graph ($\hat c_{\tilde i},F,\ell_{\tilde i},C$), and its mass ($\hat m_{\tilde i}$); the stage variable $r_n$ shifts the emphasis as the order grows. These quantities are defined next.

For each pair $(i,j)$, let $\mathcal K_{ij}$ be its index of conflict intervals. For $k \in \mathcal{K}_{ij}$, an arc-length interval $[\underline s_i^k,\overline s_i^k]$ is normalized as $Q_i^k=[\underline q_i^k,\overline q_i^k] =[\underline s_i^k/\bar s_i,\overline s_i^k/\bar s_i]$ (and likewise $Q_j^k$), making conflict regions comparable for different path lengths. Let
\begin{equation}
w_{ij}=\sum_{k\in\mathcal K_{ij}}\big(1+\delta_{ij}^k\big),\quad \delta_{ij}^k=\tfrac{1}{2}\big[(\overline q_i^k-\underline q_i^k) +(\overline q_j^k-\underline q_j^k)\big],
\label{eq:pggo_edge_weight}
\end{equation}
with $w_{ij}=0$ if $\mathcal K_{ij}=\emptyset$. In \eqref{eq:pggo_edge_weight} , $\delta_{ij}^k$ is the average normalized span. The weighted degree $c_i=\sum_{j\ne i}w_{ij}$ is the standard strength centrality~\cite{newman2018networks}; $c_{\max}=\max_i c_i$ normalizes the graph terms, $\hat c_{\tilde i}=c_{\tilde i}/c_{\max}$. The leafness $\ell_i=\max_{j\ne i}w_{ij}/c_i$ (with $\ell_i=1$ if $c_i=0$) is near one when $i$ couples to a single neighbor and smaller when its conflicts are spread out, letting \ac{PGGO} defer peripheral vehicles.

When  a candidate $\tilde i$ is compared against a group vehicle $j$, each conflict $k\in\mathcal K_{\tilde i j}$ yields, in the normalized two-vehicle space, a diagonal-blocking overlap, two precedence gaps, and a candidate-side position:
\begin{align}
d_{{\tilde i}j}^{k}&=\mathrm{ReLU}\!\big(\min(\overline q_{\tilde i}^k,\overline q_j^k)-\max(\underline q_{\tilde i}^k,\underline q_j^k)\big),\nonumber\\
g_{{\tilde i}j}^{k}&=\mathrm{ReLU}(\underline q_{\tilde i}^k-\overline q_j^k),\quad
h_{{\tilde i}j}^{k}=\mathrm{ReLU}(\underline q_j^k-\overline q_{\tilde i}^k),\quad
\xi_{{\tilde i}j}^{k}=\underline q_{\tilde i}^k.
\label{eq:PGGO_diag}
\end{align}
If $d_{{\tilde i}j}^k>0$ there are conflicts overlapping the simultaneous-progress diagonal between start node and final node in the reduced 2D-search problem. This forces the reduced path to detour; if $g_{{\tilde i}j}^k>0$ (resp.\ $h_{{\tilde i}j}^k>0$) intervals are separated, so the group (resp.\ candidate) clears first; $\xi_{{\tilde i}j}^k$ is larger for conflicts later along the candidate path. Averaging each of these quantities over $\mathcal K_{{\tilde i}j}$ gives $d_{{\tilde i}j},g_{{\tilde i}j},h_{{\tilde i}j},\xi_{{\tilde i}j}$, which are aggregated over the group with the graph weights,
\begin{equation}
D({\tilde i},\mathcal G)=\frac{\sum_{j\in\mathcal G}w_{{\tilde i}j}\,d_{{\tilde i}j}} {\sum_{j\in\mathcal G}w_{{\tilde i}j}},
\label{eq:pggo_group_diag}
\end{equation}
and $\Gamma,V,\Xi$ analogously from $g_{{\tilde i}j},h_{{\tilde i}j},\xi_{{\tilde i}j}$ (all set to zero if ${\tilde i}$ has no conflict with $\mathcal G$). Thus $D$ penalizes candidates that block simultaneous progress, $\Gamma$ and $V$ reward a clean precedence relation, and $\Xi$ mildly prefers later candidate-side conflicts.

The candidate's pressure on the unresolved set and its coupling to the group are
\begin{equation}
F(\tilde i,\mathcal U\setminus\{\tilde i\})=\frac{1}{c_{\max}}\sum_{u\in\mathcal U\setminus\{{\tilde i}\}}w_{{\tilde i}u},\quad
C(\tilde i,\mathcal G)=\frac{1}{c_{\max}}\sum_{j\in\mathcal G}w_{{\tilde i}j}.
\label{eq:pggo_graph_terms}
\end{equation}
The stage variable
\begin{equation}
r_n=\frac{n-2}{N_{\mathcal A}-2},\qquad N_{\mathcal A}=|\mathcal A|>2,\;\; n=|\mathcal G|,
\label{eq:pggo_stage}
\end{equation}
rises from $0$ to $1$ over the pass. Early on ($1-r_n$ large), $-\omega_f (\theta_F F+(1-\theta_F)\hat c_{\tilde i})$ pulls central, conflict-heavy vehicles forward while $+\omega_\ell\ell_{\tilde i}$ defers leaves, committing the conflict core first; late ($r_n$ large), $-\omega_c r_n C$ attaches the remaining vehicles where they best couple to the committed group. $F$ and $\hat c_{\tilde i}$ are blended under the single weight $\omega_f$ because early in the pass most vehicles are unresolved, so $F\approx\hat c_{\tilde i}$; $\theta_F$ sets their split. The mass term $-\omega_m\hat m_{\tilde i}$, $\hat m_{\tilde i}=m_{\tilde i}/\bar m$, acts only as a tie-breaker.

The selected ${\tilde i}^\star$ is realized by solving its reduced two-vehicle problem, advancing the shared incremental state; the analytic ranking triggers no planner call. The first pair is chosen by the same rule. Algorithm~\ref{alg:PGGO_ordering} summarizes \ac{PGGO}.

\begin{algorithm}[t]
\caption{Planner-Guided Greedy Ordering and Coordination}
\label{alg:PGGO_ordering}
\begin{algorithmic}
\Require Paths $\{\pmb{\phi}_i\}$, conflict set $\mathcal{O}$, masses $\{m_i\}$, aerodynamic factors $\{\kappa_i\}$, $v_{\max}$, $\lambda$, active set $\mathcal{A}=\{1,\ldots,N\}$
\Ensure Insertion order $\pi$, coordination trajectory $\pmb{\gamma}$

\State Build conflict graph \eqref{eq:pggo_edge_weight} and normalized conflict intervals
\State $(i_1,i_2)\leftarrow$ best seed pair in $\mathcal{A}$ by $\Psi$
\State $\mathcal{G}\leftarrow\{i_1,i_2\}$,\; $\mathcal{U}\leftarrow\mathcal{A}\setminus\mathcal{G}$,\; $\pi\leftarrow(i_1,i_2)$
\State $\mathcal{C}^{1}\leftarrow[0,\bar{s}_{i_1}]\times[0,\bar{s}_{i_2}]$
\State $\mathcal{O}^{1}\leftarrow$ project conflicts between $i_1$ and $i_2$ onto $\mathcal{C}^{1}$
\State $\pmb{\gamma}^{1}\leftarrow\textsc{EAGS}(\mathcal{C}^{1},\mathcal{O}^{1},\{m_i\},\{\kappa_i\},v_{\max},\lambda)$
\For{$n=2$ to $N-1$}
    \State ${\tilde i}^\star\leftarrow\arg\min_{{\tilde i}\in\mathcal{U}}\Psi({\tilde i}\mid\mathcal{G})$ \Comment{score \eqref{eq:pggo_selection}}
    \State $L^{n-1}\leftarrow L_\infty(\pmb{\gamma}^{n-1})$
    \State $\mathcal{C}^{n}\leftarrow[0,\bar{s}_{{\tilde i}^\star}]\times[0,L^{n-1}]$
    \State $\mathcal{O}^{n}\leftarrow$ project conflicts between ${\tilde i}^\star$ and $\mathcal{G}$ onto $\mathcal{C}^{n}$
    \State $\widehat{\pmb{\gamma}}^{n}\leftarrow
    \textsc{EAGS}(\mathcal{C}^{n},\mathcal{O}^{n},\{m_i\},\{\kappa_i\},v_{\max},\lambda,\pmb{\gamma}^{n-1})$
    \State $\pmb{\gamma}^{n}\leftarrow\textsc{Lift}(\widehat{\pmb{\gamma}}^{n},\pmb{\gamma}^{n-1})$
    \State $\pi\leftarrow(\pi,{\tilde i}^\star)$;\; $\mathcal{G}\leftarrow\mathcal{G}\cup\{{\tilde i}^\star\}$;\; $\mathcal{U}\leftarrow\mathcal{U}\setminus\{{\tilde i}^\star\}$
\EndFor
\State \Return $\pi$,\; $\pmb{\gamma}=\pmb{\gamma}^{N-1}$
\end{algorithmic}
\end{algorithm}

For each iteration $k$, 
building the weighted conflict graph and the normalized pairwise conflict quantities requires one pass over all pairwise conflict intervals. Assuming the number of conflict intervals per vehicle pair is bounded, the total number of such records is $O(N^2)$. The weighted degrees, normalized centralities, and leafness indices are also computed over all vehicle pairs, giving $O(N^2)$. Seed-pair selection ranks $O(N^2)$ candidate pairs. During the greedy pass, the number of candidate score evaluations is $\sum_{n=2}^{N-1}(N-n)=O(N^2)$. Each score evaluation is $O(1)$ because the pairwise quantities are precomputed and the sums defining $D,\Gamma,V,\Xi,F$, and $C$ are updated incrementally after each insertion. Since the \textsc{EAGS}-method dominates computational complexity, Algorithm \ref{alg:PGGO_ordering} has the same computational complexity as Algorithm~\ref{alg:incremental_state_augmented_dag}, $\mathcal{O}(Nr^3)$ that is. 

\section{Controller Design}
\label{sec:controller}

To track the reference trajectories generated by the high-level planner, we employ a \ac{NMPC} framework for each vehicle. The vehicle dynamics are modeled using the kinematic bicycle model \cite{7225830}, discretized with a sampling time $\Delta t$. The state vector is defined as $\mathbf{x}_k = [x_k, y_k, v_k, \psi_k, \delta_k]^\top$, comprising the two-dimensional Cartesian position, longitudinal velocity, heading angle, and steering angle. The control input vector is $\mathbf{u}_k = [a_k, \dot{\delta}_k]^\top$, consisting of longitudinal acceleration and steering rate.

The discrete-time system dynamics $\mathbf{x}_{k+1} = f(\mathbf{x}_k, \mathbf{u}_k)$ are given by:
\begin{equation}
    \begin{aligned}
    x_{k+1} &= x_k + v_k \cos(\psi_k) \Delta t, \\
    y_{k+1} &= y_k + v_k \sin(\psi_k) \Delta t, \\
    v_{k+1} &= v_k + a_k \Delta t, \\
    \psi_{k+1} &= \psi_k + \frac{v_k}{L} \tan(\delta_k) \Delta t, \\
    \delta_{k+1} &= \delta_k + \dot{\delta}_k \Delta t,
    \end{aligned}
\end{equation}
where $L$ denotes the vehicle wheelbase.

At each time step $t$, the NMPC solves a finite-horizon optimal control problem to minimize the deviation from the reference trajectory while satisfying physical constraints:
\begin{subequations}
\begin{align}
    \min_{\mathbf{u}_{t:t+H-1}} \quad & \sum_{k=t}^{t+H-1} \Big( \|\mathbf{x}_{k|t} - \mathbf{x}_{k|t}^{ref}\|_{\mathbf{Q}}^2 + \|\mathbf{u}_{k|t}\|_{\mathbf{R}}^2 \nonumber \\
    & \quad + \|\Delta \mathbf{u}_{k|t}\|_{\mathbf{S}}^2 \Big) + \|\mathbf{x}_{t+H|t} - \mathbf{x}_{t+H|t}^{ref}\|_{\mathbf{Q}_f}^2 \\
    \text{s.t.} \quad & \mathbf{x}_{k+1|t} = f(\mathbf{x}_{k|t}, \mathbf{u}_{k|t}), \\
    & \mathbf{x}_{k|t} \in \mathcal{X}, \quad \mathbf{u}_{k|t} \in \mathcal{U}, \\
    & \mathbf{x}_{0|t} = \mathbf{x}(t).
\end{align}
\end{subequations}

Here, $H$ is the prediction horizon. The matrices $\mathbf{Q}, \mathbf{Q}_f \succeq 0$ penalize tracking errors, while $\mathbf{R}, \mathbf{S} \succeq 0$ penalize control effort and control smoothness, respectively. The sets $\mathcal{X}$ and $\mathcal{U}$ define the admissible states and inputs, encapsulating constraints on velocity, acceleration, steering angle, and steering rate.

The reference states $\mathbf{x}_{k|t}^{ref}$ are derived from the vehicle's precomputed fixed geometric path $\pmb{\phi}(s)$. The progression along this path is computed by the reference velocity $v_{k|t}^{ref}$ obtained via the velocity mapping procedure detailed in Section~\ref{subsec:objective}. Specifically, the arc-length parameter $s$ for the reference state is advanced at each step according to $s_{k+1} = s_k + v_{k|t}^{ref} \Delta t$.

\section{Numerical Results}
\label{sec:numerical_results}
\begin{figure}[h]
    \centering
    \includegraphics[width=0.85\linewidth]{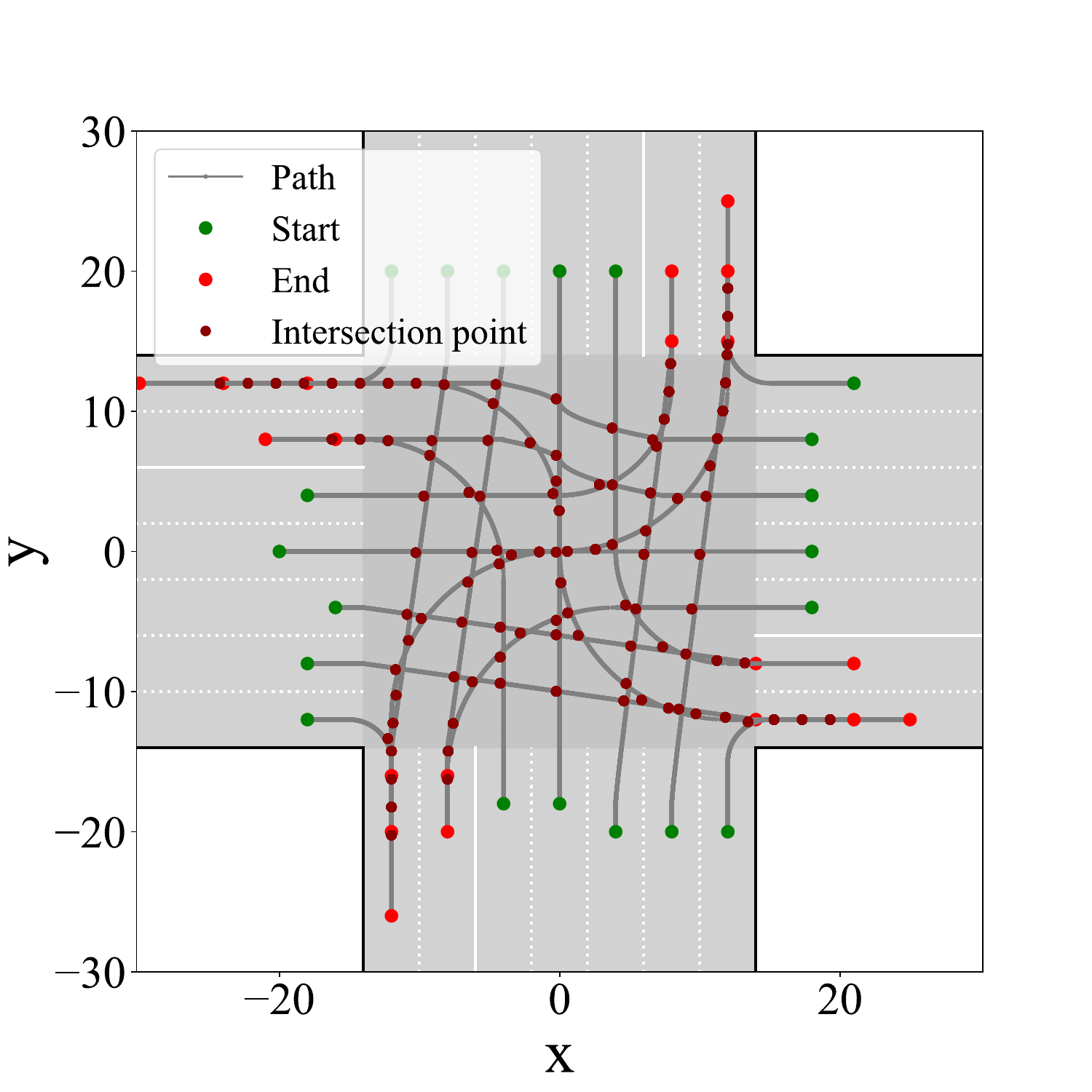}
    \caption{Intersection scenario used in the numerical evaluation. Vehicles follow predefined paths through the shared intersection area.}
    \label{fig:simulation_scenario}
\end{figure}

This section evaluates the proposed energy-aware coordination method on the intersection scenario shown in Fig.~\ref{fig:simulation_scenario}, which consists of 20 predefined paths. We consider traffic instances with 10 to 40 vehicles, generated by assigning vehicles to these paths under different vehicle-path combinations. When the number of vehicles exceeds the number of available paths, multiple vehicles are assigned to the same path and initialized at different positions along it.

The vehicles' parameters are summarized in Table~\ref{tab:vehicle_parameters}. Vehicle masses range from \SI{1100}{\kilo\gram} to \SI{2600}{\kilo\gram}, and each vehicle is assigned an individual aerodynamic drag factor $\kappa_i$, derived using its drag coefficient $C_{d,i}$ and frontal area $A_i$. 
For simplicity, in scenarios with more than 20 vehicles, for paths with more than one vehicle, those vehicles have the same mass and aerodynamic parameters. For example, in the 40-vehicle scenario there are two vehicles on each path and  vehicles 21-40 use the same parameters as vehicles 1-20, respectively.

\begin{table}[h]
\centering
\caption{Vehicle parameters used in the numerical experiments. The aerodynamic drag factor is $\kappa_i=\tfrac12\rho C_{d,i}A_i$, with air density $\rho=\SI{1.225}{\kilo\gram\per\cubic\metre}$.}
\label{tab:vehicle_parameters}
\setlength{\tabcolsep}{4pt}
\renewcommand{\arraystretch}{1.15}
\begin{tabular}{
  c
  S[table-format=4.0] S[table-format=1.2] S[table-format=1.2] S[table-format=1.3]
  @{\hspace{6pt}}c@{\hspace{2pt}}
  S[table-format=4.0] S[table-format=1.2] S[table-format=1.2] S[table-format=1.3]
}
\toprule
\textbf{Veh.} & {$m_i$} & {$C_{d,i}$} & {$A_i$} & {$\kappa_i$} &
\textbf{Veh.} & {$m_i$} & {$C_{d,i}$} & {$A_i$} & {$\kappa_i$} \\
 & {[\si{\kilo\gram}]} & & {[\si{\square\metre}]} & {[\si{\kilo\gram\per\metre}]} &
 & {[\si{\kilo\gram}]} & & {[\si{\square\metre}]} & {[\si{\kilo\gram\per\metre}]} \\
\midrule
1  & 1350 & 0.28 & 2.15 & 0.369 & 11 & 1500 & 0.28 & 2.20 & 0.377 \\
2  & 1430 & 0.29 & 2.18 & 0.387 & 12 & 1100 & 0.26 & 2.05 & 0.326 \\
3  & 1390 & 0.28 & 2.16 & 0.370 & 13 & 1580 & 0.29 & 2.22 & 0.394 \\
4  & 1760 & 0.30 & 2.32 & 0.426 & 14 & 1650 & 0.29 & 2.28 & 0.405 \\
5  & 1650 & 0.29 & 2.28 & 0.405 & 15 & 1700 & 0.30 & 2.32 & 0.426 \\
6  & 1720 & 0.30 & 2.32 & 0.426 & 16 & 1690 & 0.30 & 2.31 & 0.424 \\
7  & 1620 & 0.29 & 2.24 & 0.398 & 17 & 2150 & 0.32 & 2.55 & 0.500 \\
8  & 1580 & 0.29 & 2.22 & 0.394 & 18 & 2300 & 0.33 & 2.65 & 0.536 \\
9  & 1720 & 0.30 & 2.34 & 0.430 & 19 & 2350 & 0.34 & 2.70 & 0.562 \\
10 & 1680 & 0.30 & 2.30 & 0.423 & 20 & 2600 & 0.35 & 2.85 & 0.611 \\
\bottomrule
\end{tabular}
\end{table}

The pairwise conflict regions are constructed by placing a circular safety region of diameter \SI{12}{\metre} around each geometric path intersection. The \ac{PGGO} parameters used in all numerical experiments are summarized in Table~\ref{tab:pggo_weights}.
\begin{table}[t]
\centering
\caption{Numerical values of the \ac{PGGO} score weights.}
\label{tab:pggo_weights}
\setlength{\tabcolsep}{10pt}
\renewcommand{\arraystretch}{1.15}
\begin{tabular}{@{}c c c c@{}}
\toprule
\textbf{Weight} & \textbf{Value} & \textbf{Weight} & \textbf{Value} \\
\midrule
$\omega_d$    & $1.00$ & $\omega_f$    & $2.00$ \\
$\omega_g$    & $0.85$ & $\theta_F$    & $0.35$ \\
$\omega_v$    & $0.20$ & $\omega_\ell$ & $0.60$ \\
$\omega_\xi$  & $0.35$ & $\omega_c$    & $0.50$ \\
$\omega_m$    & $0.01$ & --            & --     \\
\bottomrule
\end{tabular}
\end{table}

All experiments were implemented in Python~3.11 and executed on a MacBook with an Apple~M4 Pro chip and \SI{24}{\giga\byte} of memory.

\subsection{Time-Energy Trade-off}

We evaluate the effect of the trade-off parameter $\lambda$ on the planned transport time and the proposed energy term, which consists of the positive kinetic-energy increment and the aerodynamic drag. For each value of $\lambda\in\{0,0.1,\ldots,1\}$, the planner is evaluated over 101 randomized vehicle-order permutations for the 20-vehicle scenario. 
$\lambda=0$ corresponds to the energy-term, while $\lambda=1$ corresponds to the time-term. The reported metrics are the planned transport time and the planned energy.
\begin{figure}[h]
\centering
\begin{tikzpicture}
\begin{axis}[
    width=0.88\columnwidth,
    height=0.64\columnwidth,
    xlabel={Mean energy term [$10^{4}$ J]},
    ylabel={Mean transport time [s]},
    label style={font=\footnotesize},
    tick label style={font=\scriptsize},
    axis line style={black,line width=0.6pt},
    tick style={gray},
    tick align=inside,
    xmin=3.75,
    xmax=5.65,
    ymin=78,
    ymax=98.5,
    ymajorgrids=false,
    xmajorgrids=false,
    colormap/viridis,
    point meta min=0,
    point meta max=1,
    clip=false,
]

\addplot[
    only marks,
    scatter,
    scatter src=explicit,
    scatter/use mapped color*={
        draw=mapped color!55!black,
        fill=mapped color
    },
    mark=*,
    mark size=3pt,
]
table[
    col sep=comma,
    x expr=\thisrow{planned_positive_kinetic_energy_J_mean}/10000,
    y=max_planned_transport_time_s_mean,
    meta=alpha
]{computed_summary_by_alpha.csv};

\addplot[
    only marks,
    mark=none,
    point meta=explicit,
    nodes near coords={
        $\pgfmathprintnumber[
            fixed,
            fixed zerofill,
            precision=1
        ]{\pgfplotspointmeta}$
    },
    every node near coord/.append style={
        font=\scriptsize,
        text=black,
        anchor=west,
        xshift=2pt
    },
]
table[
    col sep=comma,
    x expr=\thisrow{planned_positive_kinetic_energy_J_mean}/10000,
    y=max_planned_transport_time_s_mean,
    meta=alpha
]{computed_summary_by_alpha_3.csv};

\end{axis}
\end{tikzpicture}
\caption{Mean time--energy trade-off for different values of $\lambda$. Each marker represents the mean over 101 randomized vehicle-order permutations, and the marker label denotes the corresponding value of $\lambda$.}
\label{fig:time_energy_tradeoff}
\end{figure}
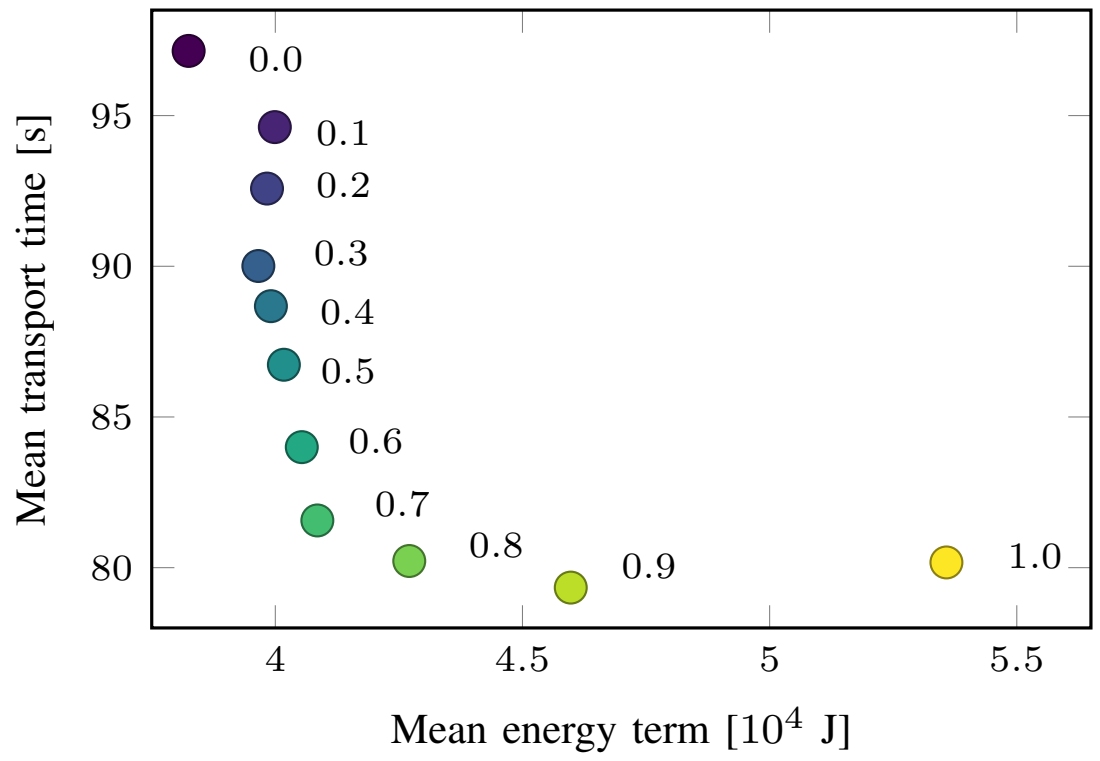

Fig.~\ref{fig:time_energy_tradeoff} shows the mean time--energy trade-off obtained by varying $\lambda$. As expected, larger values of $\lambda$ generally reduce the planned transport time by giving more weight to the time term, but this effect is not strictly monotonic for every neighboring value of $\lambda$. This is because changes in $\lambda$ may alter the selected coordination pattern and the active interaction constraints. Overall, the mean planned transport time decreases from $97.14$~s at $\lambda=0$ to $80.17$~s at $\lambda=1$, corresponding to a $17.5\%$ reduction. Over the same range, the mean energy increases from $3.88\times10^4$~J to $5.41\times10^4$~J, corresponding to a $39.5\%$ increase.

\begin{table}[t]
\centering
\caption{Selected mean results for different values of $\lambda$.}
\label{tab:lambda_selected_results}
\begin{tabular}{
c
S[table-format=2.2]
S[table-format=2.2]
S[table-format=-2.1]
S[table-format=2.1]
}
\toprule
{$\lambda$} &
{Time [s]} &
{Energy [kJ]} &
{Time change [\%]} &
{Energy change [\%]} \\
\midrule
0.0 & 97.15 & 38.797 & {--} & {--} \\
0.4 & 88.68 & 40.478 & -8.7  &  4.3 \\
0.6 & 84.00 & 41.093 & -13.5 &  5.9 \\
0.7 & 81.57 & 41.408 & -16.0 &  6.7 \\
0.8 & 80.22 & 43.265 & -17.4 & 11.5 \\
1.0 & 80.17 & 54.130 & -17.5 & 39.5 \\
\bottomrule
\end{tabular}
\end{table}

Table~\ref{tab:lambda_selected_results} shows that the improvement in transport time saturates for large values of $\lambda$, while the energy continues to increase. For example, $\lambda=0.7$ reduces the mean planned transport time by $16\%$ relative to $\lambda=0$, while increasing the energy by only $6.8\%$. In contrast, the purely time-oriented case $\lambda=1$ gives only a modest additional time reduction but causes a much larger energy increase. Therefore, intermediate values, especially $\lambda=0.6$--$0.8$, provide a favorable compromise between intersection throughput and energy demand.

\subsection{Sensitivity to Vehicle-Order Permutations}
\label{subsec:permutation_sensitivity}

\begin{figure}[h]
\centering
\resizebox{\columnwidth}{!}{%
\begin{tikzpicture}
\begin{axis}[
    width=8.0cm,
    height=5.8cm,
    view={-58}{24},
    xlabel={Permutation},
    xlabel style={font=\scriptsize, rotate=25},
    ylabel={$\lambda$},
    y dir=reverse,
    zlabel={Planned transport time [s]},
    xlabel style={font=\scriptsize},
    ylabel style={font=\scriptsize},
    zlabel style={font=\scriptsize},
    tick label style={font=\scriptsize},
    grid=major,
    xmin=0, xmax=100,
    ymin=0, ymax=1.0,
    zmin=45, zmax=130,
    xtick={0,25,50,75,100},
    ytick={0,0.2,0.4,0.6,0.8,1.0},
    colormap={energycolormap}{
        rgb255(0cm)=(255,255,190);
        rgb255(1cm)=(255,210,90);
        rgb255(2cm)=(255,140,45);
        rgb255(3cm)=(220,30,30);
        rgb255(4cm)=(150,0,40)
    },
    colorbar,
    colorbar style={
        width=1.5mm,
        ylabel={Energy [J]},
        ylabel style={font=\scriptsize},
        tick label style={font=\scriptsize},
    },
    point meta min=32281,
    point meta max=56039,
]

\addplot3[
    surf,
    shader=faceted interp,
    point meta=explicit,
    mesh/ordering=y varies,
    mesh/rows=11,
    draw=black,
    line width=0.08pt,
]
table[
    col sep=comma,
    x=seed_index,
    y=alpha,
    z=max_planned_transport_time_s,
    meta=planned_positive_kinetic_energy_J
] {alpha_sweep_results_1.csv};

\end{axis}
\end{tikzpicture}%
}
\caption{Effect of the weighting parameter $\lambda$ and vehicle-order permutation on the planned performance. The surface height indicates the maximum planned transport time, and the surface color indicates the positive kinetic-energy increment.}
\label{fig:lambda_perm_surface}
\end{figure}
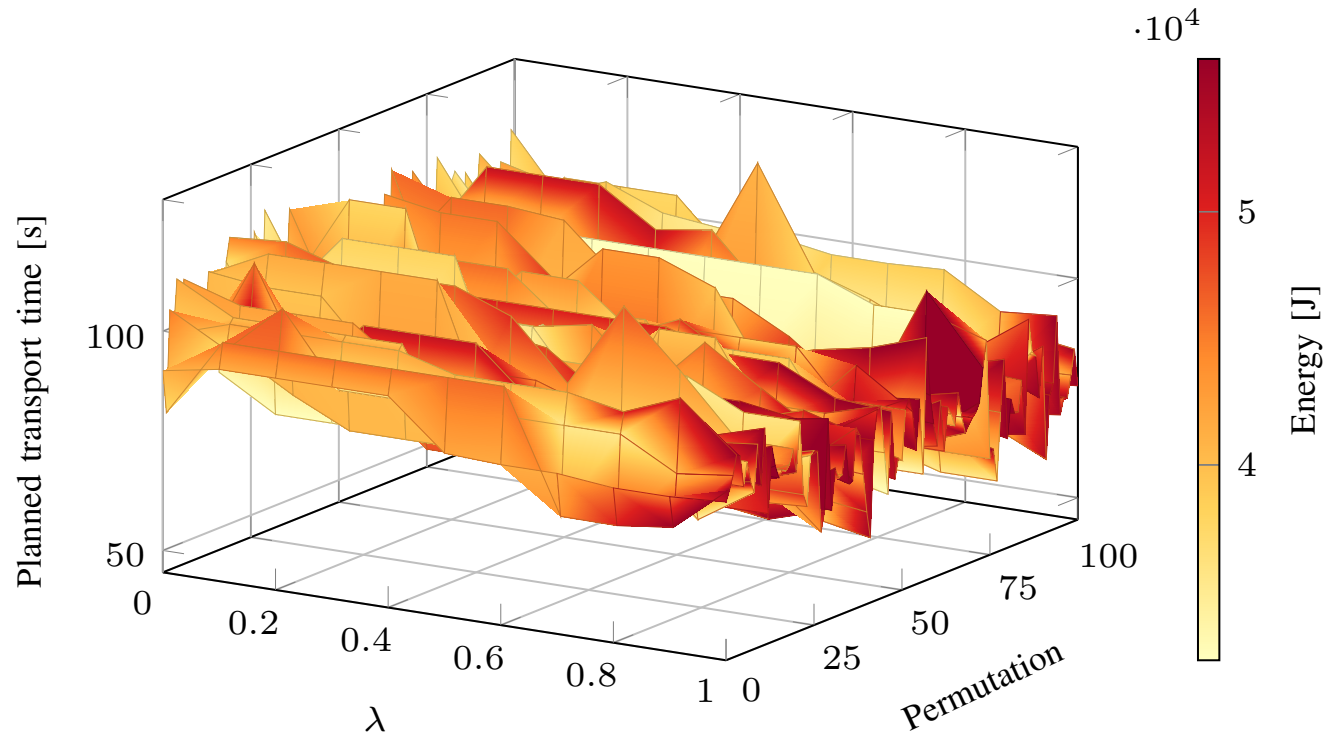

Fig.~\ref{fig:lambda_perm_surface} shows that the planner's performance depends jointly on $\lambda$ and the vehicle insertion order, motivating a principled ordering rule. To identify a practical strategy, we fix $\lambda=0.7$ and compare five approaches across scenario sizes $N\in\{10,15,20,25,30,35,40\}$. For each size the planner is evaluated over up ten vehicle-path combinations, and the random baseline is additionally averaged over $100$ random permutations per combination. The compared approaches are: (i)~\emph{Random} ordering, used as the reference distribution; (ii)~a \emph{static \ac{WCG}} rule, representative of graph-based conflict-resolution heuristics~\cite{chen2016cooperative,xu2018distributed,li2024collision}; (iii)~the proposed approach \emph{\ac{PGGO}}; and (iv)--(v)~two local-search refinements, \emph{Random+LS} and \emph{\ac{PGGO}+LS}, both obtained by a first-improvement pairwise-swap local search initialized from the random and \ac{PGGO} orders, respectively. Table~\ref{tab:pggo_scalability_summary} reports the resulting transport time, energy, and runtime.

The static conflict-graph rule provides only modest and inconsistent gains. Relative to the random mean, it changes the transport time by between $+1.0\%$ ($N=15$, a slight degradation) and $-18.1\%$ ($N=35$), and reduces energy by only $2$--$12\%$. This confirms that static pairwise conflict information alone, without the reduced-space geometry exploited by the planner, is insufficient to determine a competitive order.

\ac{PGGO} is markedly stronger. It reduces the transport time by an amount that grows with scenario size, from $4$--$17\%$ at $N\le20$ to $24\%$, $39\%$, $45\%$, and $50\%$ at $N=25,30,35$, and $40$, respectively, while its runtime is under $1$~s ($0.72$~s at $N=40$). Moreover, \ac{PGGO} improves energy; at $\lambda=0.7$ it also lowers the energy at every size, by up to about $23\%$ (e.g., $38.1$~kJ versus $49.5$~kJ at $N=20$). A single deterministic pass therefore dominates both the random and static-graph orders in solution quality at negligible cost.

Building on the \ac{PGGO} order, local search refines it further: warm-started from \ac{PGGO}, it reaches lower transport times while clearly outperforming LS started at random. \ac{PGGO}+LS attains the lowest transport time at every size $N\ge20$, reducing it relative to the random mean by $34\%$ at $N=20$ and up to $61.5\%$ at $N=40$. Its margin over Random+LS widens sharply with scenario size, the two are comparable for $N\le15$, but at $N=40$ \ac{PGGO}+LS reaches $112.6$~s against $208.6$~s for Random+LS, a $46\%$ improvement. Since both variants use identical first-improvement local search, this gap is attributable entirely to the initialization, Random+LS converges quickly to a poor local optimum from its random start, whereas the \ac{PGGO} warm start places local search in a basin yielding substantially better orders. \ac{PGGO}+LS also retains the lowest energy at almost all sizes (e.g., $92.5$~kJ versus $110.4$~kJ for random and $97.6$~kJ for Random+LS at $N=40$).

The cost of this refinement is modest and well-bounded. First-improvement keeps the search far cheaper than a full-neighborhood pass, \ac{PGGO}+LS completes in $0.79$~s at $N=10$ and at most about $64$~s at $N=40$. Although Random+LS is sometimes faster (about $28$~s at $N=40$), its lower cost reflects premature convergence to an inferior optimum rather than greater efficiency; the additional time spent by \ac{PGGO}+LS purchases a $46\%$ reduction in transport time at the largest scenario.

Taken together, these results yield a clear conclusion. \ac{PGGO} alone captures a large and increasing share of the ordering benefit deterministically and in well under a second, making it attractive for online use, while \ac{PGGO}+LS reaches the best attainable transport time and energy at a refinement cost of at most about a minute. We therefore adopt \ac{PGGO} with first-improvement local search (\ac{PGGO}+LS) as the main ordering strategy in the remaining simulations.

\begin{table}[h]
\centering
\caption{Comparison of ordering strategies at $\lambda=0.7$. For each fleet size $N$ the planner is evaluated over up to ten vehicle-path combinations; the random baseline additionally averages over $100$ permutations per combination, while the deterministic strategies (WCG, \ac{PGGO}, Random+LS, \ac{PGGO}+LS) are run once per combination. $T$ is the transport time, $E$ the planned energy, and runtime the mean wall-clock time per evaluation.}
\label{tab:pggo_scalability_summary}
\setlength{\tabcolsep}{6pt}
\renewcommand{\arraystretch}{1.15}
\begin{tabular}{@{}l ccccccc@{}}
\toprule
& \multicolumn{7}{c}{\textbf{Number of vehicles} $N$}\\
\cmidrule(l){2-8}
\textbf{Strategy} & \textbf{10} & \textbf{15} & \textbf{20}$^{\dagger}$ & \textbf{25} & \textbf{30} & \textbf{35} & \textbf{40}$^{\dagger}$\\
\midrule
\multicolumn{8}{@{}l}{\emph{Transport time $T$ [s]}}\\
Random    & 43.9 & 60.2 & 79.9 & 135.6 & 189.2 & 242.3 & 292.4\\
WCG       & 43.1 & 60.8 & 66.7 & 131.4 & 164.7 & 198.4 & 242.0\\
PGGO      & 40.8 & 57.8 & 66.1 & 103.2 & 115.9 & 134.4 & 145.7\\
Random+LS & 35.0 & 44.4 & 56.2 &  85.9 & 117.2 & 145.9 & 208.6\\
PGGO+LS   & 35.5 & 45.9 & 52.6 &  73.9 &  88.4 & 100.6 & 112.6\\
\midrule
\multicolumn{8}{@{}l}{\emph{Energy $E$ [kJ]}}\\
Random    & 25.3 & 37.0 & 49.5 & 62.7 & 76.8 & 93.8 & 110.4\\
WCG       & 24.0 & 36.2 & 43.8 & 59.2 & 74.4 & 90.0 & 104.4\\
PGGO      & 22.8 & 36.0 & 38.1 & 57.4 & 72.1 & 88.7 &  93.8\\
Random+LS & 21.2 & 32.8 & 39.9 & 57.1 & 70.1 & 85.5 &  97.6\\
PGGO+LS   & 20.3 & 32.5 & 42.7 & 54.1 & 69.3 & 84.4 &  92.5\\
\midrule
\multicolumn{8}{@{}l}{\emph{Runtime [s]}}\\
Random    & 0.01 & 0.01 & 0.03 & 0.08 & 0.16 & 0.26 &   0.37\\
WCG       & 0.01 & 0.02 & 0.03 & 0.08 & 0.15 & 0.23 &   0.33\\
PGGO      & 0.01 & 0.03 & 0.05 & 0.16 & 0.30 & 0.49 &   0.72\\
Random+LS & 0.69 & 2.88 & 2.48 & 10.74 & 25.25 & 27.31 &  27.89\\
PGGO+LS   & 0.79 & 2.94 & 3.97 & 16.69 & 38.35 & 65.91 & 64.08\\
\bottomrule
\end{tabular}
\\[3pt]
{\footnotesize $^{\dagger}$Results for $N=20$ and $N=40$ summarize one vehicle--path combination; the remaining sizes average over ten.}
\end{table}

\subsection{Comparison with Baseline Methods}
\label{subsec:baseline_comparison}

We compare the proposed planner against two representative centralized baselines. The first is the hierarchical \ac{MILP} formulation of~\cite{9611165}: its upper level fixes a globally time-optimal conflict-zone schedule through binary ordering variables, and its lower level computes the energy-optimal trajectory corresponding to that schedule. The second is the hierarchical convex framework of~\cite{9916133}, which first fixes a crossing order and then solves a \ac{SOCP} for an energy-time optimal trajectory. To ensure a fair comparison, both baselines are adapted to the present setting and use the same energy metric. All methods are evaluated over scenario sizes $N\in\{10,15,20,25,30,35,40\}$. Means and standard deviations are taken over the corresponding scenario combinations. The proposed planner is run at $\lambda\in\{0,0.7,1\}$, while the Pareto weight of the convex baseline is fixed to $0.7$. Beyond $N=25$ the convex program of~\cite{9916133} could not be solved to certified optimality on the dense conflicts, so the corresponding data points are omitted.

Fig.~\ref{fig:baseline_comparison}(a) reports the transport time. By construction, the time-optimal \ac{MILP} of~\cite{9611165} attains the lowest transport time at every size since it solves a globally time-optimal schedule. The proposed method does not minimize time globally; it relies on a deterministic insertion order (\ac{PGGO}) and an incremental 2D decomposition. Nevertheless, the trade-off weight $\lambda$ drives its transport time toward the MILP reference: at $N=20$ it falls from about $57$~s at $\lambda=0$ to $49$--$52$~s at $\lambda\in\{0.7,1\}$, within roughly $20$--$25\%$ of the global optimum. The convex baseline is markedly slower than both, yielding transport times more than double those of the proposed planner (e.g., $111$~s versus $52$~s at $N=20$, $\lambda=0.7$).

\begin{figure}[h]
\centering

\pgfplotstableread[col sep=comma]{equal_time_energy_across_combinations.csv}\equaldata

\definecolor{darkpurple}{RGB}{90,0,170}

\pgfplotsset{
  equalaxis/.style={
    width=0.98\columnwidth,
    height=0.68\columnwidth,
    xmin=8.5,
    xmax=41.5,
    ymin=3,
    ymax=110,
    xtick={10,15,20,25,30,35,40},
    ytick={20,40,60,80,100},
    xlabel={Number of vehicles},
    ylabel={Equal-time energy [kJ]},
    title style={font=\footnotesize},
    label style={font=\footnotesize},
    tick label style={font=\scriptsize},
    axis line style={black,line width=0.6pt},
    tick style={black,line width=0.6pt},
    tick align=inside,
    scaled y ticks=false,
    every axis plot/.append style={line width=1.15pt},
    error bars/error bar style={line width=0.6pt},
    error bars/error mark options={line width=0.6pt,mark size=1.4pt},
    unbounded coords=discard,
    filter discard warning=false,
    legend pos=north west,
    legend cell align={left},
    legend style={
      draw=gray!70,
      fill=white,
      fill opacity=0.92,
      text opacity=1,
      font=\footnotesize,
      row sep=1pt
    },
  },
  pzero/.style={
    normblue,
    mark=*,
    mark size=2.4pt,
    mark options={fill=normblue,draw=normblue}
  },
  psev/.style={
    normgreen,
    mark=square*,
    mark size=2.3pt,
    mark options={fill=normgreen,draw=normgreen}
  },
  pone/.style={
    normred,
    mark=triangle*,
    mark size=2.6pt,
    mark options={fill=normred,draw=normred}
  },
  cmethod/.style={
    black,
    dashed,
    mark=diamond*,
    mark size=2.5pt,
    mark options={fill=black,draw=black}
  },
  panm/.style={
    darkpurple,
    dotted,
    mark=x,
    mark size=2.9pt,
    mark options={draw=darkpurple,line width=0.8pt}
  },
  keep proposed/.style={
    x filter/.code={%
      \pgfplotstablegetelem{\coordindex}{method}\of{\equaldata}%
      \edef\mymethod{\pgfplotsretval}%
      \pgfplotstablegetelem{\coordindex}{alpha}\of{\equaldata}%
      \edef\myalpha{\pgfplotsretval}%
      \ifnum\pdfstrcmp{\mymethod}{Proposed}=0
        \ifnum\pdfstrcmp{\myalpha}{#1}=0
        \else
          \def\pgfmathresult{nan}%
        \fi
      \else
        \def\pgfmathresult{nan}%
      \fi
    },
  },
  keep method/.style={
    x filter/.code={%
      \pgfplotstablegetelem{\coordindex}{method}\of{\equaldata}%
      \edef\mymethod{\pgfplotsretval}%
      \ifnum\pdfstrcmp{\mymethod}{#1}=0
      \else
        \def\pgfmathresult{nan}%
      \fi
    },
  },
}

\begin{tikzpicture}
\begin{axis}[equalaxis]

\addplot+[pzero,keep proposed=0.0,error bars/.cd,y dir=both,y explicit]
table[
  x=N,
  y expr=\thisrow{equal_time_energy_J_mean}/1000,
  y error expr={ifthenelse(\thisrow{equal_time_energy_J_count}>1,
  \thisrow{equal_time_energy_J_std}/1000,0)}
]{\equaldata};

\addlegendentry{Proposed ($\lambda=0.0$)}

\addplot+[psev,keep proposed=0.7,error bars/.cd,y dir=both,y explicit]
table[
  x=N,
  y expr=\thisrow{equal_time_energy_J_mean}/1000,
  y error expr={ifthenelse(\thisrow{equal_time_energy_J_count}>1,
  \thisrow{equal_time_energy_J_std}/1000,0)}
]{\equaldata};
\addlegendentry{Proposed ($\lambda=0.7$)}

\addplot+[pone,keep proposed=1.0,error bars/.cd,y dir=both,y explicit]
table[
  x=N,
  y expr=\thisrow{equal_time_energy_J_mean}/1000,
  y error expr={ifthenelse(\thisrow{equal_time_energy_J_count}>1,
  \thisrow{equal_time_energy_J_std}/1000,0)}
]{\equaldata};
\addlegendentry{Proposed ($\lambda=1.0$)}

\addplot+[cmethod,keep method=Chalaki,error bars/.cd,y dir=both,y explicit]
table[
  x=N,
  y expr=\thisrow{equal_time_energy_J_mean}/1000,
  y error expr={ifthenelse(\thisrow{equal_time_energy_J_count}>1,
  \thisrow{equal_time_energy_J_std}/1000,0)}
]{\equaldata};
\addlegendentry{MILP-based}

\addplot+[panm,keep method=Pan,error bars/.cd,y dir=both,y explicit]
table[
  x=N,
  y expr=\thisrow{equal_time_energy_J_mean}/1000,
  y error expr={ifthenelse(\thisrow{equal_time_energy_J_count}>1,
  \thisrow{equal_time_energy_J_std}/1000,0)}
]{\equaldata};
\addlegendentry{Convex}

\end{axis}
\end{tikzpicture}

\caption{Energy evaluated at the equal transportation time determined by the slowest valid method for each vehicle count.}
\label{fig:equal-time-energy}
\end{figure}
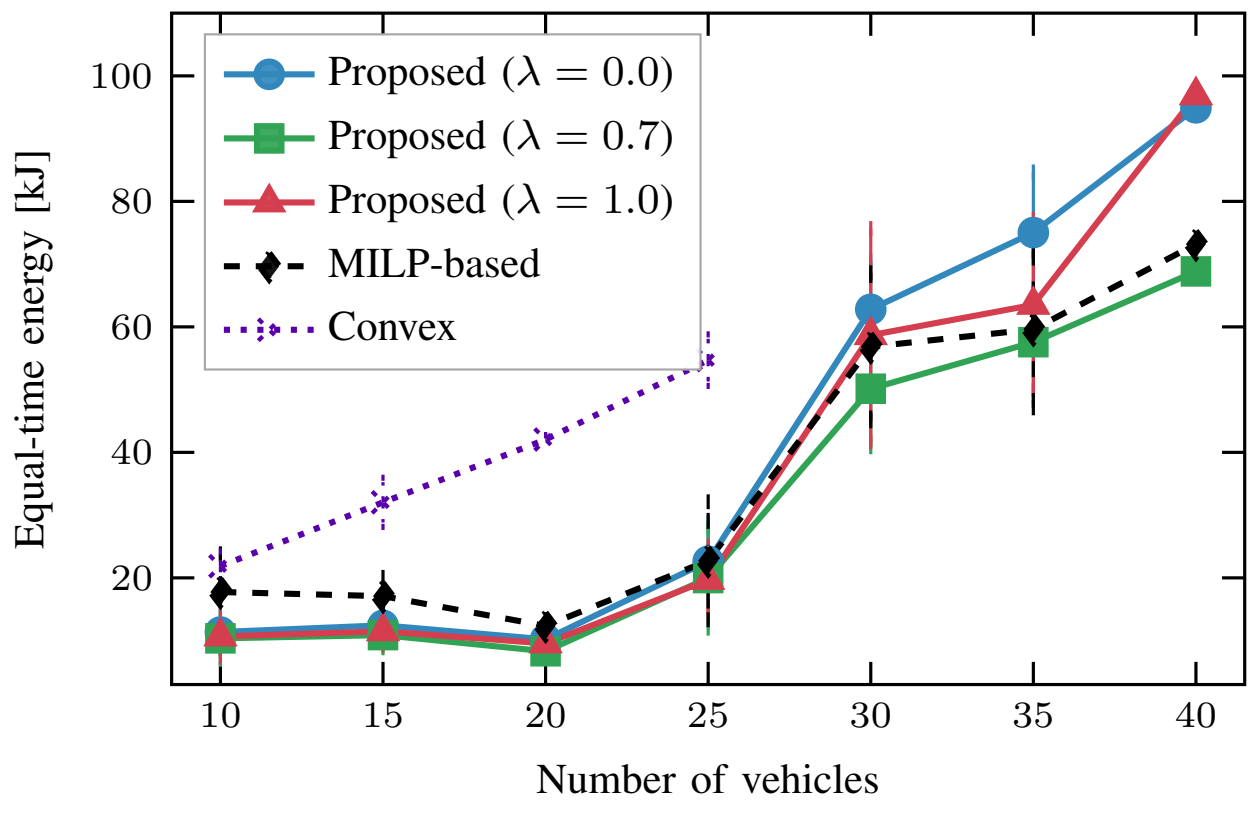

Fig.~\ref{fig:baseline_comparison}(b) shows the energy, where the proposed planner has a decisive advantage. The time-optimal MILP incurs the highest energy and grows fastest with $N$ (from about $42$~kJ at $N=10$ to $219$~kJ at $N=40$). This is because the tight time-optimal arrival windows can only be met with large accelerations. The proposed planner avoids this penalty: at $\lambda=0.7$ it lowers the energy by roughly $52$-$57\%$ relative to~\cite{9611165} for $N\ge 20$ (e.g., $38$~kJ versus $88$~kJ at $N=20$), and by about $56$--$59\%$ at $\lambda=0$. The convex baseline attains intermediate energy but is dominated by the proposed planner on both axes: at $N\ge 20$ the proposed planner at $\lambda=0$ is simultaneously faster and lower in energy (e.g., $57$~s and $38$~kJ versus $111$~s and $42$~kJ at $N=20$). To compare energy on equal terms, Fig.~\ref{fig:equal-time-energy} normalizes every method to a common transport time per scenario, set by the slowest valid method at that scenario, and re-evaluates the energy under this uniform retiming. On this common time basis the proposed planner at $\lambda=0.7$ consumes less energy than both the MILP and the convex baseline at every scenario size. 

Fig.~\ref{fig:baseline_comparison}(c) reports the computation runtime on a logarithmic scale. The MILP is inexpensive for small fleets but scales combinatorially, rising from about $0.01$~s at $N=10$ to $36$~s at $N=30$ and $600$~s at $N=40$, nearly five orders of magnitude across the tested range. The convex baseline requires several seconds even at small $N$ (about $4$-$19$~s) due to the conic solve and does not return certified-optimal solutions beyond $N=25$. The proposed planner is the only method that scales across the entire range and solves every scenario size. Its runtime grows gracefully and overtakes the MILP beyond $N\approx30$: at the largest size it is about an order of magnitude faster (roughly $9\times$ at $N=40$, $64$~s versus $600$~s).

A note on implementation is warranted for the runtime comparison. Both baselines are solved with highly optimized, compiled industrial solvers, the MILP of~\cite{9611165} with HiGHS~\cite{huangfu2018parallelizing} and the convex program of~\cite{9916133} with MOSEK~\cite{mosek}, whereas the proposed planner is implemented in Python without low-level code optimization or parallelization. The reported runtimes therefore favor the baselines on a per-operation basis, and a compiled or otherwise optimized implementation of the proposed planner would further reduce its computation time. The runtime advantage reported above is thus a conservative estimate of the proposed method's scalability.

\begin{figure*}[h]
\centering

\pgfplotstableread[col sep=comma]{full_method_comparison_across_combinations.csv}\compdata

\pgfplotsset{
  localaxis/.style={
    width=\linewidth,
    height=0.80\linewidth,
    xmin=8.5,
    xmax=41.5,
    xtick={10,15,20,25,30,35,40},
    xlabel={Number of vehicles},
    tick align=inside,
    tick style={black,line width=0.45pt},
    axis line style={black,line width=0.55pt},
    label style={font=\scriptsize},
    tick label style={font=\tiny},
    every axis plot/.append style={line width=0.95pt},
    error bars/error bar style={line width=0.45pt},
    error bars/error mark options={line width=0.45pt,mark size=0.8pt},
    unbounded coords=discard,
    filter discard warning=false,
  },
  pzero/.style={normblue,mark=*,mark size=1.35pt, mark options={fill=normblue,draw=normblue}},
  psev/.style={normgreen,mark=square*,mark size=1.35pt, mark options={fill=normblue,draw=normgreen}},
  pone/.style={normred,mark=triangle,mark size=1.50pt, mark options={fill=normblue,draw=normred}},
  cmethod/.style={black,mark=diamond,mark size=1.45pt,dashed},
  panm/.style={normpurple,mark=x,mark size=2pt,dashdotted},
  keep proposed/.style={
    x filter/.code={%
      \pgfplotstablegetelem{\coordindex}{method}\of{\compdata}%
      \edef\mymethod{\pgfplotsretval}%
      \pgfplotstablegetelem{\coordindex}{alpha}\of{\compdata}%
      \edef\myalpha{\pgfplotsretval}%
      \ifnum\pdfstrcmp{\mymethod}{Proposed}=0
        \ifnum\pdfstrcmp{\myalpha}{#1}=0
        \else
          \def\pgfmathresult{nan}%
        \fi
      \else
        \def\pgfmathresult{nan}%
      \fi
    },
  },
  keep method/.style={
    x filter/.code={%
      \pgfplotstablegetelem{\coordindex}{method}\of{\compdata}%
      \edef\mymethod{\pgfplotsretval}%
      \ifnum\pdfstrcmp{\mymethod}{#1}=0
      \else
        \def\pgfmathresult{nan}%
      \fi
    },
  },
}

\begin{tikzpicture}
\begin{axis}[
  hide axis,
  width=0pt,
  height=0pt,
  scale only axis,
  xmin=0, xmax=1,
  ymin=0, ymax=1,
  legend columns=5,
  legend cell align={left},
  legend style={
    draw=black,
    fill=white,
    font=\footnotesize,
    line width=0.5pt,
    inner sep=2pt,
    column sep=0.8em
  },
  legend image code/.code={
    \draw[#1]
    plot coordinates {(0cm,0cm) (0.35cm,0cm) (0.70cm,0cm)};
  },
  legend to name=methodlegend,
]

\addlegendimage{normblue,line width=1.0pt,mark=*,mark size=1.1pt}
\addlegendentry{Proposed ($\lambda$=0.0)}

\addlegendimage{normgreen,line width=1.0pt,mark=square*,mark size=1.1pt}
\addlegendentry{Proposed ($\lambda$=0.7)}

\addlegendimage{normred,line width=1.0pt,mark=triangle*,mark size=1.1pt}
\addlegendentry{Proposed ($\lambda$=1.0)}

\addlegendimage{black,line width=1.0pt,mark=diamond*,mark size=1.1pt,dashed}
\addlegendentry{MILP-based}

\addlegendimage{normpurple,line width=2.0pt,mark=x,mark size=1.1pt,dashdotted}
\addlegendentry{Convex}

\end{axis}
\end{tikzpicture}

\pgfplotslegendfromname{methodlegend}
\vspace{-0.35em}

\subfloat[]{%
\begin{minipage}[t]{0.33\textwidth}
\centering
\begin{tikzpicture}
\begin{axis}[
  localaxis,
  ylabel={Transportation Time [s]},
  ymin=25,
  ymax=185,
  ytick={40,80,120,160},
]
\addplot+[pzero,keep proposed=0.0,error bars/.cd,y dir=both,y explicit]
table[
  x=N,
  y=T_mean,
  y error expr={ifthenelse(\thisrow{T_count}>1,\thisrow{T_std},0)}
]{\compdata};

\addplot+[psev,keep proposed=0.7,error bars/.cd,y dir=both,y explicit]
table[
  x=N,
  y=T_mean,
  y error expr={ifthenelse(\thisrow{T_count}>1,\thisrow{T_std},0)}
]{\compdata};

\addplot+[pone,keep proposed=1.0,error bars/.cd,y dir=both,y explicit]
table[
  x=N,
  y=T_mean,
  y error expr={ifthenelse(\thisrow{T_count}>1,\thisrow{T_std},0)}
]{\compdata};

\addplot+[cmethod,keep method=Chalaki,error bars/.cd,y dir=both,y explicit]
table[
  x=N,
  y=T_mean,
  y error expr={ifthenelse(\thisrow{T_count}>1,\thisrow{T_std},0)}
]{\compdata};

\addplot+[panm,keep method=Pan,error bars/.cd,y dir=both,y explicit]
table[
  x=N,
  y=T_mean,
  y error expr={ifthenelse(\thisrow{T_count}>1,\thisrow{T_std},0)}
]{\compdata};
\end{axis}
\end{tikzpicture}
\end{minipage}}
\hfill
\subfloat[]{%
\begin{minipage}[t]{0.33\textwidth}
\centering
\begin{tikzpicture}
\begin{axis}[
  localaxis,
  ylabel={Energy [kJ]},
  ymin=0,
  ymax=230,
  ytick={50,100,150,200},
]
\addplot+[pzero,keep proposed=0.0,error bars/.cd,y dir=both,y explicit]
table[
  x=N,
  y expr=\thisrow{E_mean}/1000,
  y error expr={ifthenelse(\thisrow{E_count}>1,\thisrow{E_std}/1000,0)}
]{\compdata};

\addplot+[psev,keep proposed=0.7,error bars/.cd,y dir=both,y explicit]
table[
  x=N,
  y expr=\thisrow{E_mean}/1000,
  y error expr={ifthenelse(\thisrow{E_count}>1,\thisrow{E_std}/1000,0)}
]{\compdata};

\addplot+[pone,keep proposed=1.0,error bars/.cd,y dir=both,y explicit]
table[
  x=N,
  y expr=\thisrow{E_mean}/1000,
  y error expr={ifthenelse(\thisrow{E_count}>1,\thisrow{E_std}/1000,0)}
]{\compdata};

\addplot+[cmethod,keep method=Chalaki,error bars/.cd,y dir=both,y explicit]
table[
  x=N,
  y expr=\thisrow{E_mean}/1000,
  y error expr={ifthenelse(\thisrow{E_count}>1,\thisrow{E_std}/1000,0)}
]{\compdata};

\addplot+[panm,keep method=Pan,error bars/.cd,y dir=both,y explicit]
table[
  x=N,
  y expr=\thisrow{E_mean}/1000,
  y error expr={ifthenelse(\thisrow{E_count}>1,\thisrow{E_std}/1000,0)}
]{\compdata};
\end{axis}
\end{tikzpicture}
\end{minipage}}
\hfill
\subfloat[]{%
\begin{minipage}[t]{0.33\textwidth}
\centering
\begin{tikzpicture}
\begin{axis}[
  localaxis,
  ymode=log,
  ylabel={Runtime [s]},
  ymin=1e-2,
  ymax=1e3,
]
\addplot+[pzero,keep proposed=0.0,error bars/.cd,y dir=both,y explicit]
table[
  x=N,
  y=runtime_s_mean,
  y error plus expr={ifthenelse(\thisrow{runtime_s_count}>1,\thisrow{runtime_s_std},0)},
  y error minus expr={ifthenelse(\thisrow{runtime_s_count}>1,min(\thisrow{runtime_s_std},0.9*\thisrow{runtime_s_mean}),0)}
]{\compdata};

\addplot+[psev,keep proposed=0.7,error bars/.cd,y dir=both,y explicit]
table[
  x=N,
  y=runtime_s_mean,
  y error plus expr={ifthenelse(\thisrow{runtime_s_count}>1,\thisrow{runtime_s_std},0)},
  y error minus expr={ifthenelse(\thisrow{runtime_s_count}>1,min(\thisrow{runtime_s_std},0.9*\thisrow{runtime_s_mean}),0)}
]{\compdata};

\addplot+[pone,keep proposed=1.0,error bars/.cd,y dir=both,y explicit]
table[
  x=N,
  y=runtime_s_mean,
  y error plus expr={ifthenelse(\thisrow{runtime_s_count}>1,\thisrow{runtime_s_std},0)},
  y error minus expr={ifthenelse(\thisrow{runtime_s_count}>1,min(\thisrow{runtime_s_std},0.9*\thisrow{runtime_s_mean}),0)}
]{\compdata};

\addplot+[cmethod,keep method=Chalaki,error bars/.cd,y dir=both,y explicit]
table[
  x=N,
  y=runtime_s_mean,
  y error plus expr={ifthenelse(\thisrow{runtime_s_count}>1,\thisrow{runtime_s_std},0)},
  y error minus expr={ifthenelse(\thisrow{runtime_s_count}>1,min(\thisrow{runtime_s_std},0.9*\thisrow{runtime_s_mean}),0)}
]{\compdata};

\addplot+[panm,keep method=Pan,error bars/.cd,y dir=both,y explicit]
table[
  x=N,
  y=runtime_s_mean,
  y error plus expr={ifthenelse(\thisrow{runtime_s_count}>1,\thisrow{runtime_s_std},0)},
  y error minus expr={ifthenelse(\thisrow{runtime_s_count}>1,min(\thisrow{runtime_s_std},0.9*\thisrow{runtime_s_mean}),0)}
]{\compdata};
\end{axis}
\end{tikzpicture}
\end{minipage}}

\vspace{-0.5em}
\caption{Comparison of the proposed planner at $\lambda\in\{0,0.7,1\}$ with the adapted baselines of~\cite{9611165} (MILP-based) and~\cite{9916133} (Convex) across scenario sizes $N\in\{10,15,20,25,30,35,40\}$. (a)~Transportation time;. (b)~Energy $E$; (c)~Computation runtime (logarithmic axis). The convex baseline is omitted at $N\ge 30$, where it could not be certified optimal on these scenarios.}
\label{fig:baseline_comparison}
\end{figure*}
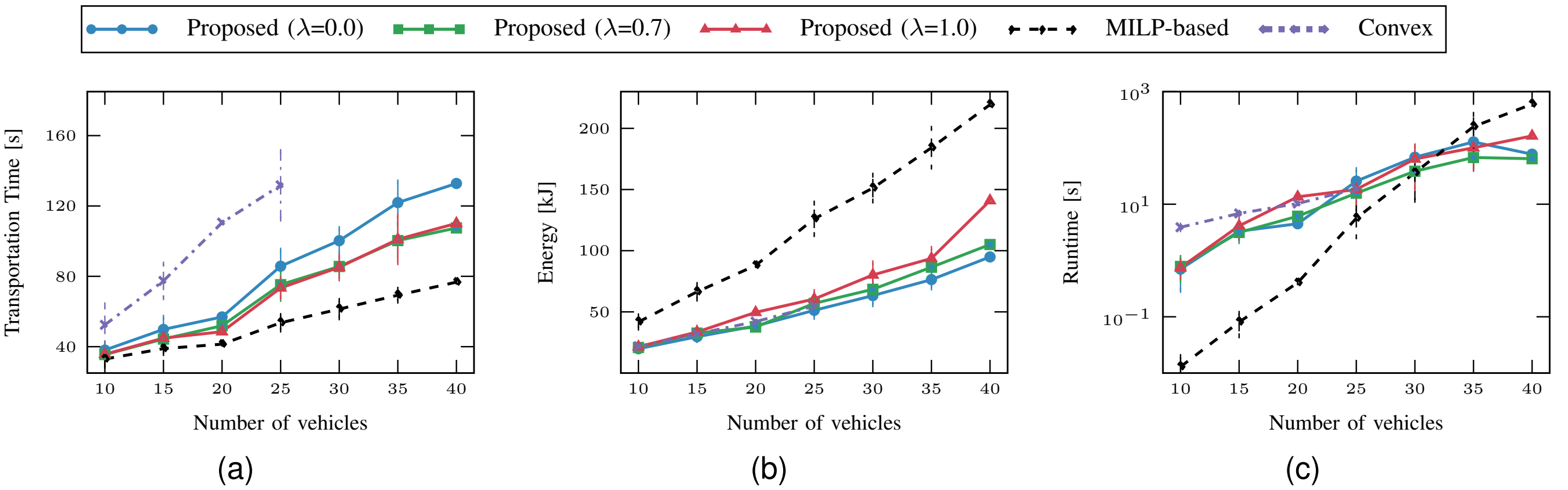

\subsection{Controller Validation}
\label{subsec:controller_validation}

To verify that the planned coordination trajectories can be executed by vehicle-level controllers, the output of the proposed planner was tracked using an \ac{NMPC} controller. The validation was performed for the 25-vehicle shared-path scenario. The controller used a sampling time of $0.1$~s and a prediction horizon of $H=30$. The vehicle wheelbase was $L=2$~m, and the imposed bounds were $v_{\max}=\SI{2}{m/s}$, $a\in[-4,4]~\si{m/s^2}$, $\delta\in[-40^\circ,40^\circ]$, and $\dot{\delta}_{\max}=\SI{0.8}{rad/s}$. The NMPC weights are reported in Table~\ref{tab:nmpc_weights}.

\begin{table}[h]
\caption{NMPC tracking weights used in the controller validation.}
    \begin{tabularx}{\linewidth}{@{}>{\raggedright\arraybackslash}X S[table-format=2.1]@{}}    \toprule
    \textbf{Description} & \textbf{Value} \\
    \midrule
    Position tracking cost                 & 50.0 \\
    Velocity tracking cost                 & 10.0 \\
    Heading tracking cost                  & 10.0 \\
    Control effort (acceleration/steering) & 0.5 \\
    Control smoothness (acceleration/steering) & 10.0 \\
    Steering smoothness   & 1.0 \\
    \midrule
    Terminal cost for final position error & 70.0 \\
    Terminal cost for final velocity error & 30.0 \\
    Terminal cost for final heading error  & 30.0 \\
    \bottomrule
  \end{tabularx}
  \label{tab:nmpc_weights}
\end{table}

Fig.~\ref{fig:velocity_tracking} shows the reference and vehicles' velocity profiles, while Fig.~\ref{fig:control_effort} reports the corresponding acceleration and steering-rate inputs. In addition, an animation of vehicle positions is provided at \url{https://youtu.be/HjpNDfHa0aU}. The results show that the planned trajectories can be tracked by the low-level controller with bounded control effort, confirming that the proposed configuration-space planner produces executable reference motions for the considered multi-vehicle scenario.

\providecommand{\nmpcValidationDataPath}{figures}

\definecolor{nmpcBlue}{RGB}{31,119,180}
\definecolor{nmpcOrange}{RGB}{255,127,14}
\definecolor{nmpcGreen}{RGB}{44,160,44}
\definecolor{nmpcRed}{RGB}{214,39,40}
\definecolor{nmpcPurple}{RGB}{148,103,189}
\definecolor{nmpcBrown}{RGB}{140,86,75}
\definecolor{nmpcPink}{RGB}{227,119,194}
\definecolor{nmpcGray}{RGB}{80,80,80}
\definecolor{nmpcOlive}{RGB}{188,189,34}
\definecolor{nmpcCyan}{RGB}{23,190,207}

\pgfplotsset{
    nmpcVeh0/.style={nmpcBlue},
    nmpcVeh1/.style={nmpcOrange},
    nmpcVeh2/.style={nmpcGreen},
    nmpcVeh3/.style={nmpcRed},
    nmpcVeh4/.style={nmpcPurple},
    nmpcVeh5/.style={nmpcBrown},
    nmpcVeh6/.style={nmpcPink},
    nmpcVeh7/.style={nmpcGray},
    nmpcVeh8/.style={nmpcOlive},
    nmpcVeh9/.style={nmpcCyan},
    nmpcVeh10/.style={nmpcBlue!65!black},
    nmpcVeh11/.style={nmpcOrange!75!black},
    nmpcVeh12/.style={nmpcGreen!70!black},
    nmpcVeh13/.style={nmpcRed!70!black},
    nmpcVeh14/.style={nmpcPurple!75!black},
    nmpcVeh15/.style={nmpcBrown!75!black},
    nmpcVeh16/.style={nmpcPink!70!black},
    nmpcVeh17/.style={nmpcGray!70!black},
    nmpcVeh18/.style={nmpcOlive!70!black},
    nmpcVeh19/.style={nmpcCyan!70!black},
    nmpcVeh20/.style={nmpcBlue!45!black},
    nmpcVeh21/.style={nmpcOrange!55!black},
    nmpcVeh22/.style={nmpcGreen!50!black},
    nmpcVeh23/.style={nmpcRed!55!black},
    nmpcVeh24/.style={nmpcPurple!55!black},
}

\begin{figure}[t]
\centering
\begin{tikzpicture}
\begin{axis}[
    width=0.98\linewidth,
    height=0.52\linewidth,
    xmin=0,
    xmax=90,
    ymin=-0.05,
    ymax=2.65,
    xlabel={Time [s]},
    ylabel={Velocity\\$[\si{m/s}]$},
    ylabel style={align=center},
    ytick={0,0.5,1.0,1.5,2.0,2.5},
    grid=none,
    tick label style={font=\scriptsize},
    label style={font=\footnotesize},
    axis line style={black,line width=0.6pt},
    tick style={black,line width=0.5pt},
    unbounded coords=discard,
    legend style={
        font=\scriptsize,
        draw=none,
        fill=none,
        legend columns=2,
        at={(0.02,0.98)},
        anchor=north west
    },
    legend cell align={left},
]

\pgfplotsinvokeforeach{0,...,24}{%
    \addplot[nmpcVeh#1, densely dashed, line width=0.45pt, opacity=0.70, forget plot]
        table[x=time_s,y=v_ref_#1,col sep=comma]
        {\nmpcValidationDataPath/nmpc_velocity_tracking_wide.csv};
}

\pgfplotsinvokeforeach{0,...,24}{%
    \addplot[nmpcVeh#1, solid, line width=0.55pt, opacity=0.90, forget plot]
        table[x=time_s,y=v_actual_#1,col sep=comma]
        {\nmpcValidationDataPath/nmpc_velocity_tracking_wide.csv};
}


\addlegendimage{black, densely dashed, line width=0.65pt}
\addlegendentry{Planned reference}

\addlegendimage{black, solid, line width=0.65pt}
\addlegendentry{NMPC tracking}


\end{axis}
\end{tikzpicture}
\caption{Velocity tracking performance of the NMPC controller for the 25-vehicle scenario validation. Dashed curves denote the planned reference velocities, and solid curves denote the velocities tracked by the NMPC controller.}
\label{fig:velocity_tracking}
\end{figure}
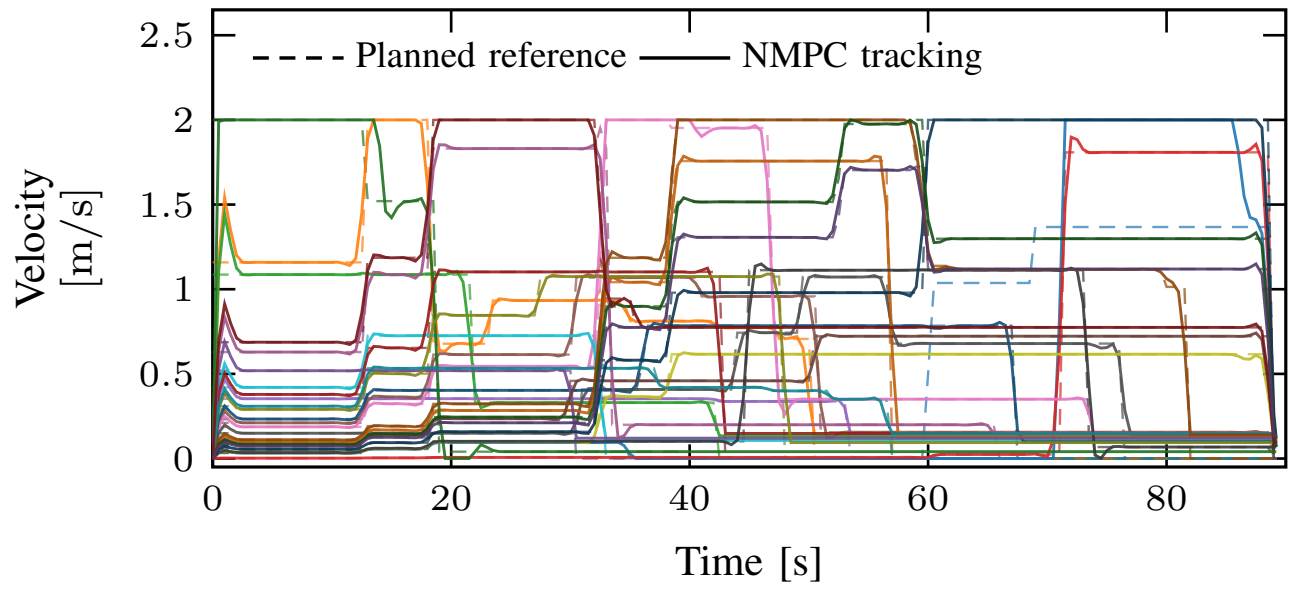
\providecommand{\nmpcValidationDataPath}{figures}

\definecolor{nmpcBlue}{RGB}{31,119,180}
\definecolor{nmpcOrange}{RGB}{255,127,14}
\definecolor{nmpcGreen}{RGB}{44,160,44}
\definecolor{nmpcRed}{RGB}{214,39,40}
\definecolor{nmpcPurple}{RGB}{148,103,189}
\definecolor{nmpcBrown}{RGB}{140,86,75}
\definecolor{nmpcPink}{RGB}{227,119,194}
\definecolor{nmpcGray}{RGB}{80,80,80}
\definecolor{nmpcOlive}{RGB}{188,189,34}
\definecolor{nmpcCyan}{RGB}{23,190,207}

\pgfplotsset{
    nmpcVeh0/.style={nmpcBlue},
    nmpcVeh1/.style={nmpcOrange},
    nmpcVeh2/.style={nmpcGreen},
    nmpcVeh3/.style={nmpcRed},
    nmpcVeh4/.style={nmpcPurple},
    nmpcVeh5/.style={nmpcBrown},
    nmpcVeh6/.style={nmpcPink},
    nmpcVeh7/.style={nmpcGray},
    nmpcVeh8/.style={nmpcOlive},
    nmpcVeh9/.style={nmpcCyan},
    nmpcVeh10/.style={nmpcBlue!65!black},
    nmpcVeh11/.style={nmpcOrange!75!black},
    nmpcVeh12/.style={nmpcGreen!70!black},
    nmpcVeh13/.style={nmpcRed!70!black},
    nmpcVeh14/.style={nmpcPurple!75!black},
    nmpcVeh15/.style={nmpcBrown!75!black},
    nmpcVeh16/.style={nmpcPink!70!black},
    nmpcVeh17/.style={nmpcGray!70!black},
    nmpcVeh18/.style={nmpcOlive!70!black},
    nmpcVeh19/.style={nmpcCyan!70!black},
    nmpcVeh20/.style={nmpcBlue!45!black},
    nmpcVeh21/.style={nmpcOrange!55!black},
    nmpcVeh22/.style={nmpcGreen!50!black},
    nmpcVeh23/.style={nmpcRed!55!black},
    nmpcVeh24/.style={nmpcPurple!55!black},
}

\begin{figure}[t]
\centering
\begin{tikzpicture}
\begin{groupplot}[
    group style={group size=1 by 2, vertical sep=0.28cm},
    width=0.98\linewidth,
    height=0.39\linewidth,
    xmin=0,
    xmax=90,
    grid=none,
    tick label style={font=\scriptsize},
    label style={font=\footnotesize},
    axis line style={black,line width=0.6pt},
    tick style={black,line width=0.5pt},
    unbounded coords=discard,
]

\nextgroupplot[
    ylabel={Acceleration\\$[\si{m/s^2}]$},
    ylabel style={align=center},
    ymin=-4.4,
    ymax=4.4,
    ytick={-4,-2,0,2,4},
    xticklabels=\empty,
    legend style={font=\scriptsize, draw=none, fill=none, at={(0.02,0.98)}, anchor=north west},
    legend cell align={left},
]

\pgfplotsinvokeforeach{0,...,24}{%
    \addplot[nmpcVeh#1, no markers, line width=0.42pt, opacity=0.85, forget plot]
        table[x=time_s,y=a_cmd_#1,col sep=comma]
        {\nmpcValidationDataPath/nmpc_control_effort_wide.csv};
}

\addlegendimage{black, line width=0.65pt}
\addlegendentry{Control input}

\addplot[nmpcRed, densely dashed, line width=0.65pt]
    coordinates {(0,4.0) (90,4.0)};
\addlegendentry{Bounds}

\addplot[nmpcRed, densely dashed, line width=0.65pt, forget plot]
    coordinates {(0,-4.0) (90,-4.0)};

\nextgroupplot[
    xlabel={Time [s]},
    ylabel={Steering rate\\$[\si{rad/s}]$},
    ylabel style={align=center},
    ymin=-0.9,
    ymax=0.9,
    ytick={-0.8,-0.4,0,0.4,0.8},
]

\pgfplotsinvokeforeach{0,...,24}{%
    \addplot[nmpcVeh#1, no markers, line width=0.42pt, opacity=0.85, forget plot]
        table[x=time_s,y=steer_rate_#1,col sep=comma]
        {\nmpcValidationDataPath/nmpc_control_effort_wide.csv};
}

\addplot[nmpcRed, densely dashed, line width=0.65pt, forget plot]
    coordinates {(0,0.8) (90,0.8)};

\addplot[nmpcRed, densely dashed, line width=0.65pt, forget plot]
    coordinates {(0,-0.8) (90,-0.8)};

\end{groupplot}
\end{tikzpicture}
\caption{NMPC control inputs for the 25-vehicle scenario validation.}
\label{fig:control_effort}
\end{figure}
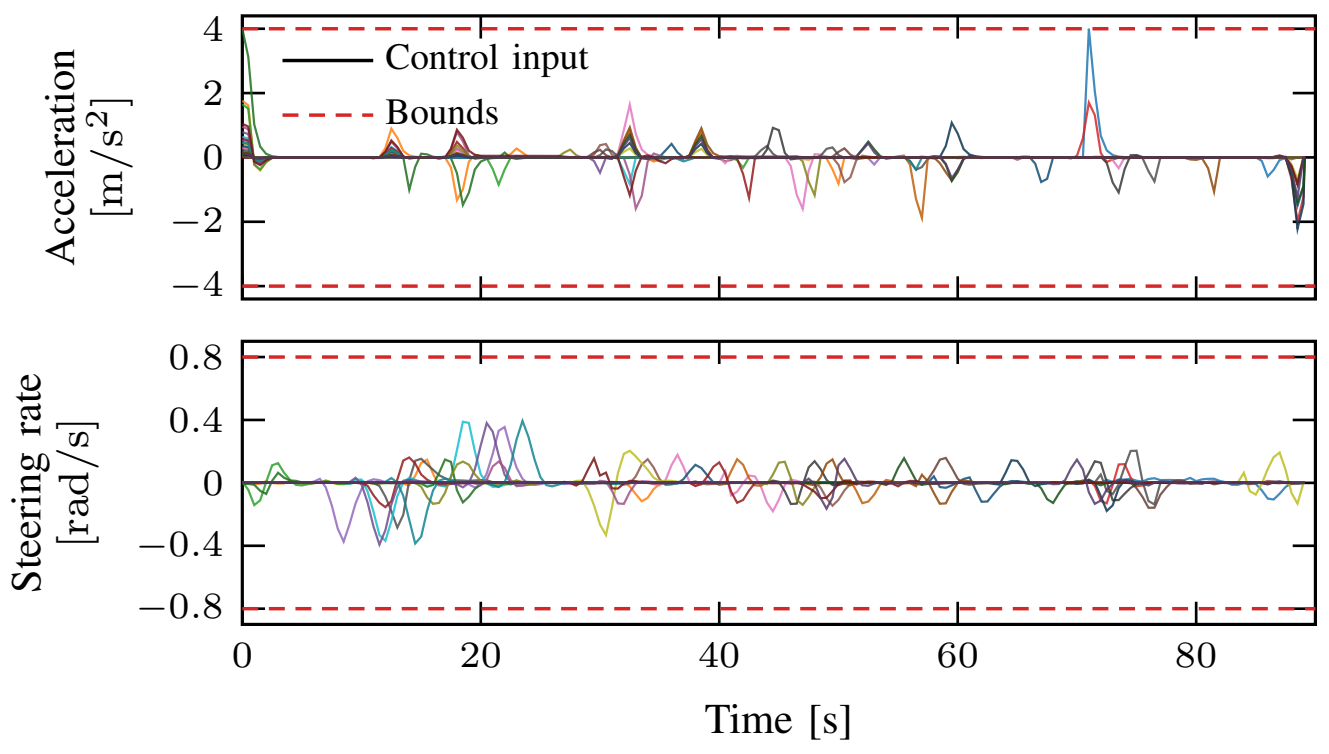

\section{Conclusion}\label{sec:conclusion}

This paper presented a scalable framework addressing joint optimization of travel time and energy consumption in multi-vehicle coordination. By integrating a physically motivated expression of energy consumption into an augmented configuration-space graph-search planner, the proposed approach enables scalable energy-aware trajectory planning. Simulations at dense intelligent intersections demonstrate that the method computes solutions significantly faster than optimization-based baselines while achieving substantial reductions in energy consumption for equal travel times. Future work will extend the framework to mixed traffic involving human-driven vehicles and explicitly account for communication and connectivity.

\bibliographystyle{IEEEtran}
\bibliography{ref}

@article{akbari20262,
  title={2-D Decompositions of High-Dimensional Configurations for Efficient Multi-Vehicle Coordination at Intelligent Intersections},
  author={Akbari, Amirreza and Thunberg, Johan},
  journal={IEEE Transactions on Intelligent Transportation Systems},
  year={2026},
  publisher={IEEE}
}

@article{lozano1979algorithm,
  author  = {Lozano-P{\'e}rez, Tom{\'a}s and Wesley, Michael A.},
  title   = {An Algorithm for Planning Collision-Free Paths Among Polyhedral Obstacles},
  journal = {Communications of the ACM},
  volume  = {22},
  number  = {10},
  pages   = {560--570},
  year    = {1979}
}

@book{latombe1991robot,
  author    = {Latombe, Jean-Claude},
  title     = {Robot Motion Planning},
  publisher = {Kluwer Academic Publishers},
  year      = {1991}
}

@book{cormen2009introduction,
  author    = {Cormen, Thomas H. and Leiserson, Charles E. and Rivest, Ronald L. and Stein, Clifford},
  title     = {Introduction to Algorithms},
  edition   = {3rd},
  publisher = {MIT Press},
  year      = {2009}
}

@article{winter2002modeling,
  author  = {Winter, Stephan},
  title   = {Modeling Costs of Turns in Route Planning},
  journal = {GeoInformatica},
  volume  = {6},
  number  = {4},
  pages   = {345--361},
  year    = {2002},
  doi     = {10.1023/A:1020853410145}
}

@article{lewis2020algorithms,
  title={Algorithms for finding shortest paths in networks with vertex transfer penalties},
  author={Lewis, Rhyd},
  journal={Algorithms},
  volume={13},
  number={11},
  pages={269},
  year={2020},
  publisher={MDPI}
}

@article{ahn2002estimating,
  title={Estimating vehicle fuel consumption and emissions based on instantaneous speed and acceleration levels},
  author={Ahn, Kyoungho and Rakha, Hesham and Trani, Antonio and Van Aerde, Michel},
  journal={Journal of transportation engineering},
  volume={128},
  number={2},
  pages={182--190},
  year={2002},
  publisher={American Society of Civil Engineers}
}

@article{RAKHA200449,
title = {Development of VT-Micro model for estimating hot stabilized light duty vehicle and truck emissions},
journal = {Transportation Research Part D: Transport and Environment},
volume = {9},
number = {1},
pages = {49-74},
year = {2004},
issn = {1361-9209},
doi = {https://doi.org/10.1016/S1361-9209(03)00054-3},
url = {https://www.sciencedirect.com/science/article/pii/S1361920903000543},
author = {Hesham Rakha and Kyoungho Ahn and Antonio Trani}
}

@article{huang2018eco,
  title={Eco-driving technology for sustainable road transport: A review},
  author={Huang, Yuhan and Ng, Elvin CY and Zhou, John L and Surawski, Nic C and Chan, Edward FC and Hong, Guang},
  journal={Renewable and sustainable energy reviews},
  volume={93},
  pages={596--609},
  year={2018},
  publisher={Elsevier}
}

@article{BARTH2009400,
title = {Energy and emissions impacts of a freeway-based dynamic eco-driving system},
journal = {Transportation Research Part D: Transport and Environment},
volume = {14},
number = {6},
pages = {400-410},
year = {2009},
note = {The interaction of environmental and traffic safety policies},
issn = {1361-9209},
doi = {https://doi.org/10.1016/j.trd.2009.01.004},
url = {https://www.sciencedirect.com/science/article/pii/S1361920909000121},
author = {Matthew Barth and Kanok Boriboonsomsin}
}

@book{golub2013matrix,
  title={Matrix computations},
  author={Golub, Gene H and Van Loan, Charles F},
  year={2013},
  publisher={JHU press}
}

@article{chen2016cooperative,
  author  = {Lei Chen and Cristofer Englund},
  title   = {Cooperative Intersection Management: A Survey},
  journal = {IEEE Transactions on Intelligent Transportation Systems},
  volume  = {17},
  number  = {2},
  pages   = {570--586},
  year    = {2016},
  doi     = {10.1109/TITS.2015.2471812}
}

@article{xu2018distributed,
  author  = {B. Xu and S. E. Li and Y. Bian and S. Li and X. J. Ban and J. Wang and K. Li},
  title   = {Distributed Conflict-Free Cooperation for Multiple Connected Vehicles at Unsignalized Intersections},
  journal = {Transportation Research Part C: Emerging Technologies},
  volume  = {93},
  pages   = {322--334},
  year    = {2018}
}

@article{li2024collision,
  author  = {Yang Li and Min Liu and Qinghai Yang and Zhong Shen and Weihua Wu},
  title   = {Collision-Free Autonomous Scheduling at Unsignalized Intersection Using Conflict Graph Tree Search},
  journal = {IEEE Internet of Things Journal},
  volume  = {11},
  number  = {8},
  pages   = {14563--14578},
  year    = {2024}
}

@ARTICLE{9611165,
  author={Chalaki, Behdad and Malikopoulos, Andreas A.},
  journal={IEEE Transactions on Intelligent Transportation Systems}, 
  title={Time-Optimal Coordination for Connected and Automated Vehicles at Adjacent Intersections}, 
  year={2022},
  volume={23},
  number={8},
  pages={13330-13345},
  doi={10.1109/TITS.2021.3123479}}

@ARTICLE{9916133,
  author={Pan, Xiao and Chen, Boli and Timotheou, Stelios and Evangelou, Simos A.},
  journal={IEEE Transactions on Intelligent Transportation Systems}, 
  title={A Convex Optimal Control Framework for Autonomous Vehicle Intersection Crossing}, 
  year={2023},
  volume={24},
  number={1},
  pages={163-177},
  doi={10.1109/TITS.2022.3211272}}

@ARTICLE{11435876,
  author={Hadjigeorgiou, Andreas and Timotheou, Stelios},
  journal={IEEE Access}, 
  title={Energy Consumption Optimization for Autonomous Vehicles via Positive Control Input Minimization}, 
  year={2026},
  volume={14},
  number={},
  pages={47651-47665},
  doi={10.1109/ACCESS.2026.3674459}}

@book{newman2018networks,
  title={Networks},
  author={Newman, Mark},
  year={2018},
  publisher={Oxford university press}
}

@INPROCEEDINGS{7225830,
  author={Kong, Jason and Pfeiffer, Mark and Schildbach, Georg and Borrelli, Francesco},
  booktitle={2015 IEEE Intelligent Vehicles Symposium (IV)}, 
  title={Kinematic and dynamic vehicle models for autonomous driving control design}, 
  year={2015},
  volume={},
  number={},
  pages={1094-1099},
  doi={10.1109/IVS.2015.7225830}}

@article{luo2023computationally,
  title={A computationally efficient bi-level coordination framework for CAVs at unsignalized intersections},
  author={Luo, Jiping and Zhang, Tingting and Zhang, Qinyu},
  journal={IEEE Transactions on Vehicular Technology},
  volume={73},
  number={2},
  pages={1868--1878},
  year={2023},
  publisher={IEEE}
}

@article{gholamhosseinian2022comprehensive,
  title={A comprehensive survey on cooperative intersection management for heterogeneous connected vehicles},
  author={Gholamhosseinian, Ashkan and Seitz, Jochen},
  journal={IEEE Access},
  volume={10},
  pages={7937--7972},
  year={2022},
  publisher={IEEE}
}

@article{hafner2021survey,
  title={A survey on cooperative architectures and maneuvers for connected and automated vehicles},
  author={H{\"a}fner, Bernhard and Bajpai, Vaibhav and Ott, J{\"o}rg and Schmitt, Georg A},
  journal={IEEE Communications Surveys \& Tutorials},
  volume={24},
  number={1},
  pages={380--403},
  year={2021},
  publisher={IEEE}
}

@article{fayazi2018mixed,
  title={Mixed-integer linear programming for optimal scheduling of autonomous vehicle intersection crossing},
  author={Fayazi, Seyed Alireza and Vahidi, Ardalan},
  journal={IEEE Transactions on Intelligent Vehicles},
  volume={3},
  number={3},
  pages={287--299},
  year={2018},
  publisher={IEEE}
}

@article{liu2020high,
  title={High throughput vehicle coordination strategies at road intersections},
  author={Liu, Changhao and Zhang, Yixiao and Zhang, Tingting and Wu, Xuanli and Gao, Lin and Zhang, Qinyu},
  journal={IEEE Transactions on Vehicular Technology},
  volume={69},
  number={12},
  pages={14341--14354},
  year={2020},
  publisher={IEEE}
}

@article{huangfu2018parallelizing,
  author  = {Huangfu, Qi and Hall, J. A. J.},
  title   = {Parallelizing the dual revised simplex method},
  journal = {Mathematical Programming Computation},
  volume  = {10},
  number  = {1},
  pages   = {119--142},
  year    = {2018},
  doi     = {10.1007/s12532-017-0130-5}
}

@manual{mosek,
   author = "MOSEK ApS",
   title = "The MOSEK Python Fusion API manual. Version 11.0.",
   year = 2025,
   url = "https://docs.mosek.com/latest/pythonfusion/index.html"
 }

@article{riostorres2016survey,
  author={Rios-Torres, Jackeline and Malikopoulos, Andreas A.},
  title={A Survey on the Coordination of Connected and Automated Vehicles at Intersections and Merging at Highway On-Ramps},
  journal={IEEE Transactions on Intelligent Transportation Systems},
  volume={18}, number={5}, pages={1066--1077}, year={2017}}

@article{malikopoulos2018decentralized,
  author={Malikopoulos, Andreas A. and Cassandras, Christos G. and Zhang, Yue J.},
  title={A Decentralized Energy-Optimal Control Framework for Connected Automated Vehicles at Signal-Free Intersections},
  journal={Automatica}, volume={93}, pages={244--256}, year={2018}}

@article{hult2018optimal,
  author={Hult, Robert and Zanon, Mario and Gros, S\'{e}bastien and Falcone, Paolo},
  title={Optimal Coordination of Automated Vehicles at Intersections: Theory and Experiments},
  journal={IEEE Transactions on Control Systems Technology},
  volume={27}, number={6}, pages={2510--2525}, year={2019}}

@article{hadjigeorgiou2023realtime,
  author={Hadjigeorgiou, Andreas and Timotheou, Stelios},
  title={Real-Time Optimization of Fuel-Consumption and Travel-Time of {CAVs} for Cooperative Intersection Crossing},
  journal={IEEE Transactions on Intelligent Vehicles},
  volume={8}, number={1}, pages={313--329}, year={2023}}

@article{han2019fundamentals,
  author={Han, Jihun and Vahidi, Ardalan and Sciarretta, Antonio},
  title={Fundamentals of Energy Efficient Driving for Combustion Engine and Electric Vehicles: An Optimal Control Perspective},
  journal={Automatica}, volume={103}, pages={558--572}, year={2019}}

@article{pan2023hierarchical,
  author={Pan, Xiao and Chen, Boli and Dai, Li and Timotheou, Stelios and Evangelou, Simos A.},
  title={A Hierarchical Robust Control Strategy for Decentralized Signal-Free Intersection Management},
  journal={IEEE Transactions on Control Systems Technology},
  volume={31}, number={5}, pages={2011--2026}, year={2023}}

@article{naderi2024lanefree,
  author={Naderi, Mehran and Papageorgiou, Markos and Troullinos, Dimitrios and Karafyllis, Iasson and Papamichail, Ioannis},
  title={Controlling Automated Vehicles on Large Lane-Free Roundabouts},
  journal={IEEE Transactions on Intelligent Vehicles},
  volume={9}, number={1}, pages={3061--3074}, year={2024}}

@article{le2024stochastic,
  author={Le, Viet-Anh and Chalaki, Behdad and Tzortzoglou, Filippos N. and Malikopoulos, Andreas A.},
  title={Stochastic Time-Optimal Trajectory Planning for Connected and Automated Vehicles in Mixed-Traffic Merging Scenarios},
  journal={IEEE Transactions on Control Systems Technology}, year={2024}}

@ARTICLE{7987071,

  author={Ahn, Heejin and Del Vecchio, Domitilla},

  journal={IEEE Transactions on Automatic Control}, 

  title={Safety Verification and Control for Collision Avoidance at Road Intersections}, 

  year={2018},

  volume={63},

  number={3},

  pages={630-642},

  doi={10.1109/TAC.2017.2729661}}

@book{lavalle2006planning,
  title={Planning algorithms},
  author={LaValle, Steven M},
  year={2006},
  publisher={Cambridge university press}
}
\end{document}